\documentclass[manuscript,screen]
{acmart}
\AtBeginDocument{%
  \providecommand\BibTeX{{%
    \normalfont B\kern-0.5em{\scshape i\kern-0.25em b}\kern-0.8em\TeX}}}

\copyrightyear{2023}
\acmYear{2023}
\setcopyright{acmlicensed}\acmConference[CHI '23]{Proceedings of the 2023 CHI Conference on Human Factors in Computing Systems}{April 23--28, 2023}{Hamburg, Germany}
\acmBooktitle{Proceedings of the 2023 CHI Conference on Human Factors in Computing Systems (CHI '23), April 23--28, 2023, Hamburg, Germany}
\acmPrice{15.00}
\acmDOI{10.1145/3544548.3580648}
\acmISBN{978-1-4503-9421-5/23/04}

\usepackage{graphicx}

\begin{document}

\title[Survivor-Centered Transformative Justice]{Survivor-Centered Transformative Justice: An Approach to Designing Alongside Domestic Violence Stakeholders in US Muslim Communities}
 

\author{Hawra Rabaan}
\email{hrabaan@iu.edu}
\orcid{0000-0002-0800-2946}
\author{Lynn Dombrowski}
\email{lsdombro@iupui.edu}
\orcid{0000-0002-9831-3229}

\affiliation{
 \institution{Indiana University-Purdue University Indianapolis}
  \streetaddress{P.O. Box 1212}
\city{Indianapolis}
\state{Indiana}
 \country{USA} 
  }


\begin{abstract}
While domestic violence (DV) is prevalent in all socioeconomic settings, identity highly impacts how one experiences and recovers from abuse. This work examines US-based Muslim women's challenges when seeking help and healing from domestic violence. Through participatory interviews with 23 participants within the DV ecosystem, we find that victim-survivors' autonomy is compromised throughout the abuse, within their immediate communities, and when involving the criminal justice system.
To address such harms, we adapt a survivor-centered transformative justice (SCTJ) approach, a framework to discern individual and systemic harm, to understand how to design \textit{alongside} victim-survivors, and to focus on victim-survivors' autonomy. We explain under what conditions an SCTJ approach may be productive for designers. We use insights from our interviews to highlight intervention areas for reducing harm, repairing harm, and promoting healing for victim-survivors. Lastly, we offer guidelines to design for harm reduction, accountability, and systemic change.
\end{abstract}



\maketitle

\section{Introduction}\label{intro}

Gender-based violence (GBV), defined as "any type of violence directed at an individual based on their gender," "is rooted in gender inequality, the abuse of power, and harmful norm"\cite{wilcenter}. Domestic violence (DV) is the most prevalent form of GBV~\cite{wilcenter}. We use the term domestic violence (DV) to encompass intimate partner violence and family violence, including violence inflicted by intimate partners, parents, siblings, children, extended family members, in-laws, community members, and community institutions, all of which "can be directly and actively involved in dynamics of abuse"\cite{kim} and enable abuse. Domestic violence is not solely a result of individual violent acts, but rather ought to be considered a systemic, public health issue due to its widespread, harmful effects \cite{chrisler2006violence, stubbs2002domestic,afrouz2020seeking}.
Domestic violence's ramifications on victim-survivors include social isolation, deprivation of education and work, financial dependence, and trauma, which reflect on their opportunities to leave the abuser, rebuild their lives after they flee, and their decisions during and after the abuse \cite{van2022posttraumatic, lee2009intimate}. Further, the victim-survivor's identity and socioeconomic background shape the way they experience and respond to DV, where victim-survivors from marginalized communities deal with a myriad of social, systemic, and legal intricacies \cite{goodmark2004law, coker, genfive, nocella2011overview, drost2015restorative, armatta2018ending}, which we portray in our work. 

Our work focuses on the American Muslim community, who we identify as people who live in the United States and self-identify, culturally or religiously, with any Islamic sect \cite{sedgwick2000sects}. We chose to focus on American Muslims for the following reasons: within HCI and the social sciences, very few studies focused on the prevalence and interventions of DV among the American Muslim population~\cite{hammer2019peaceful,oyewuwo2016american}. Muslims are an at-risk population of prejudice and systemic discrimination in the US, suffering historic Islamophobia starting in the early 20th century and rising ever since~\cite{hammer2019peaceful}. Nearly 75\% of Muslim Americans either know or have experienced acts of discrimination against Muslims following September 11, 2001, and recently, hate crimes have been amplified by hate speech promoted by politicians and the Muslim ban~\cite{oyewuwo2016american, lajevardi2018old, ali2017impact, abu2018muslim, lipka2017muslims, abadi2015majority}. Such factors resulted in a complex relationship between the American Muslim community and law enforcement. This complicated relationship directly reflects on Muslim victim-survivors, given the tight coupling of DV and the criminal justice system, forcing them to refrain from state services for protection and aid~\cite{hammer2019peaceful,oyewuwo2016american}, further contributing to victim-survivors' challenges in their response to abusive situations. 

 Our study finds that, for US-based Muslim women, seeking DV services is often not straightforward. Muslim victim-survivors commonly turn to multiple avenues to seek support and intervention, including faith leaders (\textit{i.e.,} imams), community members, and mainstream service providers -- each with their own limitations and strengths in providing care for Muslim populations. Mainstream service refers to government-funded services, non-profits, shelters, counselors, police units, and lawyers working in the DV space. Mainstream services typically follow a secular and individualistic approach to countering domestic violence as opposed to culturally-based services tailored toward the Muslim population's religious and cultural nuances. Service and support providers often fail to center women's needs in their approach \cite{sokoloff2005domestic, oyewuwo2020black, goodmark2004law, kim2021transformative}. Centering Muslim women in the DV context is to have resources that fit cultural, religious, and gendered expectations, and respect survivors' autonomy (\textit{i.e.,} recognizing and respecting the victim-survivor's capacity for self-determination \cite{verkerk1999care}) in social, cultural, and legal processes related to domestic violence, which is not the norm \cite{kim2021transformative,chatterjee, sokoloff2005domestic, goodmark2004law, oyewuwo2020black}. Currently, mainstream services are not always well suited for Muslim survivors, and some have considerable risk (\textit{e.g.,} revictimization), complicating the navigation of services and assistance \cite{kim2021transformative, oyewuwo2016american, oyewuwo2020black}. Given the notable limitations of mainstream services, victim-survivors commonly turn to Muslim faith leaders for religious and spiritual guidance, mediation, and support \cite{hammer2019peaceful, oyewuwo2016american, oyewuwo2020black}, who often have key limitations in supporting DV victim-survivors. Imams are rarely adequately trained to deal with DV's intricacies, further harming the victim-survivor by failing to take prompt action, projecting their personal views (\textit{e.g.,} tolerating abuse may be seen as a form of virtuous patience), and in some cases, siding with the abuser ~\cite{hammer2019peaceful,afrouz2020seeking}. 
 Across such services and resources, concerns persist about a lack of centering victims, supporting survivors, and taking a structural perspective, which helps us understand why abusers abuse and what perpetuates abuse. We turn to restorative justice (RJ) and transformative justice (TJ) theories to help explain and elucidate these issues while centering the victim-survivor's agency. 
 In our inquiry, we use survivor-centered transformative justice (SCTJ) as an analytic lens for our data. We describe SCTJ as an approach that uses complementary RJ and TJ principles in supporting victim-survivors' agency, encouraging community support and accountability, and accounting for and protecting against social and structural harm (described in more detail in \hyperlink{SCTJ}{Section 2.2}). We demonstrate how combining SCTJ and design allows researchers and designers to 1) account for the inequalities endured by victim-survivors, 2) address underlying causes of complex social issues by working alongside disadvantaged groups and within existing structures, and 3) understand how community members define and prefer to work towards justice~\cite{genfive, nocella2011overview, mia, Asad}.
Finally, we use insights from our participatory interviews to highlight possible intervention areas for reducing harm, repairing harm, and promoting healing within different scales (\textit{e.g.,} micro, mezzo, and macro) of relationships for DV victim-survivors.

We use the term victim-survivor to refer to the person who faced abuse due to the importance of both terms, in acknowledgment that one term does not adequately identify the experiences of the person who has undergone domestic abuse. The term(s) used depends on the person's preference and the time of the abuse. Typically, \textit{victim} is used for individuals who recently experienced DV, when discussing abusive events, or when interacting with the criminal justice system. Whereas \textit{survivor} is used for individuals who are going through the healing process \cite{SAKI, forceVS}. In restorative and transformative justice literature, the description "a person who has done harm" is used to imply abuse is a behavior that can be restored and unlearned rather than an innate trait \cite{mia, kim2021transformative}. Though we stand with its premise, throughout our work, we use the term abuser, perpetrator, offender, and person who has harmed interchangeably.

Our research contributes an understanding of the forms of abuse victim-survivors undergo and the challenges they face when seeking help formally or informally within US-based Muslim communities. We introduce and provide ways to adapt a survivor-centered transformative justice approach to addressing individual and collective harm, and implications to design for harm reduction, accountability, and systemic change. We do not wish to portray nor claim American Muslims as a monolithic group. Rather, we look at their commonly shared experiences around social and systemic injustices. In this way, our findings may be transferable to other religious US-based minorities and Muslims living in similar Western settings.  

 


\section{Background and Literature Review}\label{litrev}

 In this section, we explain the range of domestic violence services in the United States, specifically highlighting the context and experiences of American Muslims. We will highlight different approaches to justice, explicitly comparing transformative to restorative justice, explaining the applicability and challenges of different approaches when considering the context of American Muslims experiencing DV, and introduce the \textit{survivor-centered transformative justice} approach, which blends key aspects of transformative and restorative justice to center victim-survivors. Lastly, we explore how the HCI literature has engaged with justice models within DV-related scholarship. Our contribution lies in demonstrating how various justice-oriented theories may center a victim-survivor's agency in the design process and introduce a tailored approach that serves their needs and strives to protect against potential systemic and social harms when dealing with abuse and its consequences. 
 
\subsection{Domestic Violence Laws, Services, and American Muslims}\label{USmuslims}
Domestic violence is a widespread, pernicious problem. In the United States, nearly 20 people are physically abused per minute on average, and one in four women commonly between the ages of 18-24, while one in seven women have been injured by their partners \cite{NCDV}. Intimate partner violence makes up 15\% of all violent crimes. A study conducted in 2011 with 801 American Muslims revealed approximately 53\% reported facing forms of family abuse, and 31\% experienced IPV in their lifetime \cite{RN195}. The most prevalent form of abuse is emotional at 45\%, followed by verbal at 41\%, physical at 31\%, financial at 16\%, and sexual abuse at 15\% \cite{RN195}. 
Despite the widespread nature of domestic violence, DV was only recently criminalized in 1994 by the US government, through implementing the Violence Against Women Act (VAWA). VAWA allows for legal protection of DV victims along with federally funded services to aid survivors in their departure from the offender~\cite{USDOJ}.

Numerous services are provided to DV victim-survivors by governmental entities and private organizations. In our research, we refer to government-funded services, non-profits, shelters, counselors, police units, and lawyers working in the DV space as mainstream DV service providers; such services typically follow a secular and individualistic approach to countering domestic violence. Many factors contribute to challenges of mainstream DV service providers in serving niche populations such as American Muslims, including the massive burden on mainstream service providers to serve the diverse US, the complexity of DV as an individual, social, and structural problem, and the recent criminalization of DV \cite{oyewuwo2016american, Abugideiri, kim2021transformative}. 

There are several key obstacles to American Muslim women seeking help and reporting abuse. One is the diverse demographic; the majority are immigrants of ethnicities from over 50 countries \cite{ghafournia2017muslim} resulting in language and cultural understanding barriers, limited knowledge of available DV and legal services, and the perceived high financial and social costs to pursuing services \cite{oyewuwo2016american}. Additionally, Islamophobia, racism, and the over-policing of people of color and minorities are all obstacles for Muslim Americans seeking help. Roots of hostility against Muslim Americans goes long before 9/11, starting in the early 20th century~\cite{pew2021,hammer2019peaceful} and has been on the rise since the September 11 attacks, amplified by hate speech promoted by political figures and discriminatory proclamations \cite{musban, abadi2015majority, lipka2017muslims,pew2021}. Muslim men are portrayed as violent and targeted as a threat to the country, while Muslim women are viewed as victims of their religion and Muslim men~\cite{afnan2022aunties, hammer2019peaceful, gokariksel2017intersectional}.

This tense atmosphere affected Muslims nationally, in addition to a general religious misunderstanding and unfamiliarity with religious needs by non-Muslims (\textit{e.g.,} Halal food, praying space, multiple meanings of modesty), leads to the discomfort and isolation of Muslim women and delaying intervention until they are desperate \cite{ghafournia2017muslim,oyewuwo2016american}. To overcome service-seeking obstacles for Muslim women, Islamic organizations dedicated specialized efforts to countering domestic abuse in the US to tackle the gaps in service provision and address the root causes of domestic abuse~\cite{hammer2019peaceful}. The work of specialized organizations includes direct service agencies providing educational, occupational, and housing services for Muslim women and children (\textit{e.g.,} Muslimat An-Nisa\footnote{Muslimat Al-Nisaa Shelter. (2018). Retrieved from https://mnisaashelter.org/programs/}, the Domestic Harmony Foundation (DHF)\footnote{Domestic Harmony Foundation. (n.d.). Retrieved from https://dhfny.org/ }, SNS\footnote{SNS. (2013). Retrieved November 25, 2019, from http://www.sistersnurturingsisters.org/index.html.}), 
building empathy and competency nationwide through training social workers and community and religious leaders (\textit{e.g.,} the Peaceful Family Project\footnote{Peaceful families project: Working toward preventing all types of abuse in Muslim families. (n.d.). Retrieved from https://www.peacefulfamilies.org/}), and legal organizations aiming to debunk myths and empower women to gain their rights in the United States (\textit{e.g.,} Karama\footnote{Karamah: Muslim Women Lawyers for Human Rights . (n.d.). Retrieved September 15, 2022, from https://karamah.org/ }). In our study, we examine issues in mainstream DV service provision as experienced by the American Muslim community and highlight the challenges and opportunities for design to intervene within DV as experienced by Muslim women.

\subsection{Alternative Theories of Justice and Domestic Violence: Defining Survivor-Centered Transformative Justice }\label{SCTJ}\hypertarget{SCTJ}
The rise of alternative forms of justice, notably restorative and transformative justice, attempts to replace or reduce reliance on a criminal justice system (CJS) that is heavily punitive rather than rehabilitative towards offenders, and disproportionately and deleteriously affects people of color and marginalized populations  \cite{kim2021transformative, TJhistory, drost2015restorative,stubbs2002domestic}. 
Restorative justice (RJ) is an approach to addressing crime by involving the affected community members in identifying and addressing harm and promoting healing collectively away from the CJS \cite{zehr2015little, Xiao2022, kim2021transformative}. Transformative justice (TJ) attends to harm inflicted on community members while concurrently addressing broader systemic and social factors that enable harm to occur~\cite{nocella2011overview, Asad, TJsharifa, coker, TJCS, TJRabaan, genfive,chordiaTJ}. While RJ and TJ do not have singular agreed-upon definitions, both shift harm from the individual to the collective in responses and repair and are mainly distinct in their relationship with the criminal legal system. Restorative justice is intended as a replacement for the CJS and overlooks structural causes of harm \cite{kim2021transformative}, which critics see as leaving the CJS and its harmful issues intact (\textit{e.g.,} racism, Islamophobia) \cite{coker, kim2021transformative}. Whereas transformative justice actively criticizes systemic injustices contributing to harm and encourages the average person to participate in its liberatory vision of justice, and "renders the intervention of violence and its prevention as an everyday democratic act"\cite{kim2021transformative}. 


Applying such theories differs based on context, and conflicts surface when using them for domestic abuse. Restorative Justice has been mainly practiced and tested on incident-based harm between victim(s) and offender(s) who are not well-known to each other (\textit{e.g.,} juvenile justice). In contrast, domestic violence is about repeated and intentional tactics, acts of control, and "strategies that attempt to implement gender ideologies"\cite{stubbs2002domestic} between intimates or family members. Treating DV as incidents of discrete violent episodes narrows DV to individual harm, ignoring the social and political context and dismissing the abuser's deeply held attitudes and beliefs around gender and power \cite{stubbs2002domestic, coker}. 
Transformative justice theorists view the offender as a conscious choice maker, and community norms as non-neutral and rather shaped by structures, potentially being harmful for DV victim-survivors~\cite{coker}. For example, research has shown that "people often fail to condemn non-violent controlling behaviours such as threats to take children, control of money, isolation of the woman, and extreme jealousy"~\cite{coker}.
Another issue identified within RJ literature when dealing with domestic gendered violence, is its emphasis on ending the offense by focusing on apology and reparation through RJ's \textit{conferencing} process, which involves the victim-survivor, abuser, their supporters, and a facilitator to develop a harm-repairing resolution \cite{coker}. Such focus dismisses knowledge that abusers commonly use apology to manipulate their victims and others \cite{stubbs2002domestic}. Conferencing may also expose the victim-survivor to further harm through interacting with the abuser or the pressure to comply with what the community wants rather than her needs~\cite{islam2018challenges, coker}. 
Further, the act of \textit{restoration} assumes there existed a state of non-harm for both the victim and abuser, which in DV within marginalized communities is not always the case (\textit{e.g., }victim-survivors and perpetrators face systemic harm and trauma requiring comprehensive healing) \cite{coker}. Though TJ processes and practices are currently less defined compared to RJ, TJ aims to transform and create communities that support women's autonomy, and allows the opportunity 
to acknowledge the influencing systems of oppression, however, does not excuse violent behavior \cite{coker}. In our case, racism and Islamophobia are rampant systematically, resulting in economic and social disadvantages among other consequences. Faith leaders who are front-line responders to DV cases within their communities can use their religious status to support patriarchal practices and justify abusive behaviors~\cite{islam2018challenges, hammer2019peaceful}. Thus, to transform the Muslim-American woman's experience, collective trauma needs to be healed and religious institutions must develop proper training and prioritize resources to effectively deal with domestic violence within their community.


Implementing alternative justice models to different contexts is complex; idealized views of RJ strictly involving community members and practices away from the CJS, or of TJ in its pure abolitionist form without giving grace to the current changes and structural benefits for victim-survivors can result in rigidity, and the decentering of victims \cite{kim2021transformative}. 
A dilemma lies in balancing acting upon DV as a civil rights issue while avoiding state control of 
marginalized individuals and communities on the one hand and relying on community praxis for victim-survivors' justice while protecting them against norms and members enabling the abuse \cite{coker, kim2021transformative, stubbs2002domestic}.
As a result, RJ has been limitedly used in the US for DV, rather, an evolving process of justice has been promoted combining both approaches where RJ's community engagement values have been integrated within some TJ movements \cite{kim2021transformative}. 
Thus, we introduce and adapt a\textit{ Survivor-Centered Transformative Justice} approach, which uses complementary tenets from restorative and transformative justice. These tenets include: 1) to center victim-survivor's agency within current structures \cite{coker, stubbs2002domestic}; 2) to encourage taking into account RJ's insights on the importance of social networks in restraining offences and caring for victim-survivors and their needs \cite{coker, Xiao2022}; 3) lastly, to address DV as a structural, social, and political issue which requires a multi-dimensional collective approach, using a critical and rehabilitative lens to control harm and build better futures, and capitalizing on the benefits of the victim-survivor's choices \cite{kim2021transformative, stubbs2002domestic, coker}.
Throughout our paper, for simplicity we use the acronym SCTJ to refer to survivor-centered transformative justice. 

\subsection{HCI and Social Justice}
The field of HCI, specifically in its third wave, emphasizes inclusivity and human values as emerging technologies become inseparable from our daily lives \cite{filimowicz2018new}.
At the forefront of values are justice and equality. HCI scholars have proposed justice theories to tackle social platforms moderation \cite{Xiao2022, schoenebeck2021drawing, im2021yes}, community-based collaborations~\cite{Asad}, computing participation of Black and Latina girls \cite{TJCS}, and child sexual abuse \cite{TJsharifa}.

 Theoretically relevant to our work is the Prefigurative Design framework, which Asad presents as "a tool to articulate a vision with community partners to better identify opportunities to leverage existing justice work through research intervention" \cite{Asad}. Prefigurative design embraces principles of research justice and transformative justice, focusing on enabling healing within the community. Whereas empirically relevant is the work of scholars Sultana et al. (2022) that presents TJ values through design by prototyping a child sexual assault reporting tool in Bangladesh, where they aimed to include the community and work on changing root causes of child abuse through educating guardians and bringing awareness to the community~\cite{TJsharifa}. Sultana et al. argue that social change through sociotechnical system design needs to adopt local morals and values, and interventions to include the broader community rather than burdening the victim-survivors of abuse. Thus far, alternative justice models in HCI have primarily focused on issues in the public sphere (\textit{e.g.,} sexual harassment on social media platforms). In our work, we extend HCI scholarship by contending with assault done privately and repeatedly, making it all the more difficult to address. We contribute to HCI by providing a theoretical and design guide to using SCTJ in the context of harm within an underrepresented community. We focus on highlighting the systemic and social conditions enabling abuse, the abusive behavior, and consequences of abuse on the victim-survivor and the Muslim community in the US, and provide interventions that alter, repair, and reduce harm within different scales of relationships.  

\subsection{HCI and Gender-based Violence}
Gender-based violence has been studied broadly within HCI~\cite{ndjibu2017gender}. Specifically, domestic violence (DV) or Intimate Partner Violence (IPV) has been addressed by multiple scholars from the perspectives of: the role of technology in the US IPV ecosystem~\cite{freed2017digital}, DV service provision and prevention~\cite{ndjibu2017gender, Bellini2020, mechanisms, bellini2019mapping}, technology exploitation for abuse \cite{freed2018, tseng2020tools, freed2019, Leitao}, agency practices addressing DV~\cite{rab21}, designing within patriarchal systems \cite{RN5}, for survivors' security, privacy, and safety \cite{RN63,dieterle2015designing, havron2019clinical, tseng2022care}, and life-repair after DV~\cite{RN67}. Scholars have shown their concern around conventional technological and non-technological approaches that tend to burden the survivor with coping and healing while excluding abusers from the process~\cite{Leitao,mechanisms}. An exception is the recent work of Bellini et al. (2020), situated in a third-sector organization in the UK, where they aim to challenge and change the perpetrator's behavior through design. Their work is focused on preventing abusive behavior by encouraging perpetrators to be self-aware, acknowledge the extent of harm done, provide support, and respect authority~\cite{mechanisms}. The gist of Bellini et al. (2020)'s work is around designing for responsibility, where technology is leveraged to encourage the development of non-abusive practices, which complies with the values of TJ and RJ~\cite{coker,nocella2011overview,islam2018challenges}.


Scholarship of HCI and DV remains in its infancy and primarily within Western contexts. In this work we make the following contributions: we expand the DV ecosystem to include faith leaders and their liaisons, male allies, and community members, examine DV dynamics within the Muslim minority in the US, identify American-Muslims' challenges they face when seeking DV services, and illustrate technological interventions inspired by social work practices and social justice theories and models. 


\section{Methods}\label{method}
\textbf{Methods and approach:} We purposely sought out multiple opposing viewpoints through virtual semi-structured participatory design interviews (more details are provided in the \hyperlink{interviews}{Participatory Design Interviews section}). Sessions were held virtually due to the COVID-19 pandemic. Our participants included survivors, social workers, and mosque representatives to help situate Muslim women's DV experiences and their challenges in the US by the stakeholder groups directly involved~\cite{costanza2018design}. We focused on recruiting social workers who worked at cultural-based nonprofits that attended to Muslim needs among other ethnicities and religions.
We interviewed 23 participants; 14 were service providers who worked with culturally-based DV countering agencies, five Muslim-identifying DV survivors, four imams (\textit{i.e.,} faith leaders), and two mosque liaisons, who are mosque designated persons of contact specifically for female attendees. The service providers' experiences included CEOs, senior program directors, and case managers, some with over 30 years of experience. One imam was a woman and at least four of our participants identified as survivors while serving as service providers, community leaders, or advocates (further details listed in \hyperlink{Table 1}{Table 1}). Our participants' diverse backgrounds and the plurality of experiences allowed for a deeper understanding of survivors' personal, organizational, and systemic challenges.

\textbf{Recruitment:}
The recruitment process started with collecting an exhaustive list of national organizations focused on the Muslim community in the US. Overall, over 80 points of contact were either directly emailed, messaged through their websites, or reached out to on their social media pages. The second recruitment strategy was to post on relevant social media groups. Lastly, the first author tapped into her personal and professional network, and the participants' referrals for snowball sampling \cite{snowball}.

Survivors were compensated with a \$20 gift card as a form of incentive, appreciation for their time, and for sharing their personal experiences. Recruiting and interviewing survivors required flexibility; for example, one survivor was only able to talk on the phone while walking outside the house after her husband left for work and her kids were at school. We stopped recruiting service providers after reaching saturation \cite{fusch2015we}, and for imams and survivors after finding resonant themes across the conversations.

\begin{table}\hypertarget{Table 1}{}
\caption{\label{tab:Table 1}Participants' Demographics.}
$\begin{array}{ | l | l | l | l |}
\hline
	\textbf{\textit{Session\#}} &	\textbf{\textit{Participant \ ID}} & \textit{\textbf{Sex}} & \textit{\textbf{Occupation-Experience} }\\ \hline
	1 & P01 & F & Director / Survivor  \\ \hline
	2 & P02,P03 & F,F & Counselors \\ \hline
	3 & P04 & F & Executive \ Director   \\ \hline
	4 & P05 & F & Program \ Manager \\ \hline
	5 & P06 & F & Survivor \\ \hline
	6 & P07 & F & Case \ Manager \\ \hline
	7 & P08 & F & DV \ Advocate \\ \hline
	8 & P09 & F & Nonprofit \ CEO \\ \hline
	9 & P10 & F & Nonprofit \ President  \\ \hline
	10 & P11 & F & Senior \ Director \\ \hline
	11 & P12 & F & Director  \\ \hline
	12 & P13 & F & Social \ Worker \\ \hline
	13 & P14 & F & Survivor  \\ \hline
	14 & P15, P16 &  M, F  & Imam, Mosque \ Liaison\\ \hline
	15 & P17  &  M  & Imam \\ \hline
	16 & P18 &  M  & Imam \\ \hline
	17 & P19 & F & Mosque \ Liaison/Survivor/Advocacy \ Trainer \\ \hline
	18 & P20 & F & Survivor \\ \hline
	19 & P21 & F & Survivor | Advocacy | Trainee  \\ \hline
	20 & P22 & F & Survivor \\ \hline
	21 & P23 & F & Imam | Survivor \\ \hline
\end{array}$
\end{table}

\textbf{Participatory Design Interviews: }\hypertarget{interviews} The interviews were all conducted in English by the first author, who identifies as a Muslim woman. The interview protocols differed based on our participants' expertise and experience; we had three protocols slightly modified for DV professionals, survivors, and community leaders. The overarching focus of the study, as communicated to the participants and approved by the IRB committee, was to identify the gaps and challenges Muslim DV survivors faced within US-based mainstream service provision.
Each interview was divided into two parts. The first part empirically investigated the roles participants took on in the community, their experience with DV, the key challenges survivors faced when responding to DV, the use of technology in their daily practice related to DV, and the challenges they faced using technology. 
The second part of the interview was participatory-based, continuing to build on answers from the first half. Interviewees were prompted to envision "magical" social, ubiquitous, or mobile technologies that would help overcome the challenges they mentioned in the first half of the interview. Participants were encouraged to use a pen and paper to express their suggestions visually and given the option to think aloud or write on a piece of paper. Most participants were most comfortable vocally expressing their interventions, some wrote lists of ideas, and only one participant used a tablet to sketch a design, emulating an exercise she did during a course on technology for innovative solutions as part of her Master's in Social Work. 

\textbf{Ethical Considerations:} Participants were provided with the interview protocol and study information sheet before the interviews. The PI stated the option to stop the interview and refrain from answering any questions when needed verbally before starting the interview and in written form in the study information sheet. The interviewer prompted participants on their help-seeking process details and challenges rather than the details of the abuse. In the event that participants persisted in sharing abusive details, the interviewer listened attentively and offered genuine empathizing words. None of our participants were in active danger to the best of our knowledge.
The first author shared relevant resources verbally with participants when prompted during the interviews and later compiled and delivered a document with tool-kits for imams created by DV organizations and DV-related resources and mobile applications to imam participants who showed interest in acquiring such knowledge and as a gratitude gesture for their participation. 
To handle the difficulty of the topic, the interviewer established some personal care practices such as limiting interviews within a time frame and decompressing by taking nature walks after each session. 

Lastly, only the participants' preferred first names were shared to ensure confidentiality during interviews. Audio recordings were collected based on the participants' permission and were deleted upon transcription. Data were anonymized in storage and write-ups and saved securely on the university server, where only the first author had access to the audio files and transcriptions. 

\textbf{Data Analysis:} The data analysis process happened over an extended period. The first round of analysis was conducted in 2021 and focused on identifying the challenges Muslim women face in mainstream services, barriers within their immediate communities, and how to address the challenges through technical and non-technical interventions. The data was inductively and deductively analyzed~\cite{strauss1997grounded}. The first round of analysis was done separately for every user group (\textit{i.e.,} survivors, service provider, and faith leaders), then cross-coded to identify the common themes across all three groups. The concepts of transformative and restorative justice were raised explicitly by two of our participants as alternative models to the current flaws in the criminal justice system, which led to considering alternative justice models in our second round of analysis done in 2022. We formed the survivor-centered transformative justice conditions and the different scales of interventions applicable to those conditions deductively from the literature and inductively from our data. IRB approval was obtained prior to conducting the study.

Though we do not claim to generalize our findings to all Muslim women, there are transferable aspects in their legal and cultural details to at-risk, immigrant, and minority populations in the US.

\textbf{Limitations: }
A limitation could be that we did not speak to mainstream service providers (\textit{e.g.,} shelter staff, police members) to understand their points of view, nor to abusers to grasp the injustices they face in the system and society.
Further, tensions arise when applying activist theories such as transformative justice to complicated social contexts; communities can act as enablers of abuse and supporters of victims-survivors, faith leaders can help in the mediation process and healing from abuse and revictimize survivors, systems provide laws and resources that protect harmed and vulnerable individuals while inflicting harm on marginalized communities. Striking a balance between supporting the parties involved in DV dynamics while protecting victim-survivors is strenuous, however, these are \textit{justice-making processes} \cite{coker} by gradually improving the multi-layered approach of centering survivors' autonomy and decision-making in their help-seeking and healing processes while encouraging accountability and transforming communities. 
\section{Why should designers use Survivor-Centered Transformative Justice?}\label{TJconditions}
\hypertarget{TJconditions}{}
In this section, we answer the question "why should designers use SCTJ?" from our data and supported by literature. We demonstrate how combining SCTJ and design allows researchers to account for the inequalities endured by victim-survivors and community members, address underlying causes of difficult social issues by working alongside disadvantaged groups and within complex and complicit systems, and understand how community members define and prefer to work towards justice. 
Though we illustrate SCTJ within our context, we intend for this section to provide a transferable framing for other contexts where designers wish to understand the scale and scope of long-standing social problems, work within existing structures while building new infrastructures, and finally, center harmed individuals by respecting their choices. 
\subsection{To Understand Social Inequality and Complexities}
Transformative justice was initially created by and for communities experiencing systemic oppression, where a group of people is intentionally disadvantaged based on their identity, such as class, race, gender, or religious beliefs~\cite{mia}. In the case of Muslim Americans, they are a minority population who have been historically surveilled by the state \cite{said2010terrorist, afnan2022aunties}, targeted by law enforcement \cite{said2010terrorist}, and face religious, immigration status, and gendered discrimination \cite[Ch.2, p38-39]{hammer2019peaceful}\cite{oyewuwo2016american}, resulting in collective trauma \cite{ali2017impact} and a fraught relationship with state resources and services \cite{oyewuwo2016american,afrouz2020seeking}.
Muslims may hesitate to interact with the criminal justice system tied to DV services due to the complicated relationship between law enforcement and Muslims in the US~\cite{said2010terrorist,Aaronson}. This relationship is complicated for several key reasons including incidents of questionable charges, deportations, and detentions~\cite{shiekh2011detained,pulse}, community infiltration and surveillance~\cite{said2010terrorist, afnan2022aunties}, and inadequate handling of hate crimes~\cite{ProPublica}. Also, law enforcement's sometimes lax outreach efforts often exacerbate mistrust among local Muslim communities.

The widespread discrimination in governmental entities against Muslims (\textit{i.e.,} Islamophobia) \cite{freeland2001treatment, ACLU, islamo} and, in return, their lack of trust in the system was echoed by our advocates, survivors, and imams alike. Islamophobia is described as a fear, prejudice, or intolerance towards Islam or Muslims, "characterised by suspicion, deep-rooted prejudice, ignorance, and, in some cases, physical and verbal harassment"\cite{allen2020towards}. Specifically related to Muslim women, participants highlighted the issue of gendered Islamophobia (\textit{i.e.,} prejudice towards Muslim women). 
Such prejudice was reported by participants during interactions with school counselors, shelters, and public services. Viewing Muslim victim-survivors as less worthy of care impedes survivors from service seeking and burdens them to tolerate the abuse.

Another identity-based vulnerability for immigrant Muslim women is their immigration status, which abusive partners often use to control their partners (\textit{e.g.,} using deportation threats) \cite{kasturirangan2004impact,rana2012addressing}. Further, immigrants and refugees are commonly exposed to trauma caused by involuntary relocation, preimmigration circumstances, and the migration process (\textit{e.g.,} living in grim refugee camps)~\cite{rees2007domestic, lee2009intimate}. In the case of victim-survivors, trauma resulting from the abuse is compounded by earlier trauma (\textit{e.g.,} forced migration)~\cite{rees2007domestic}. 

Complex inequality requires multidimensional interventions, which survivor-centered transformative justice approaches offer by addressing the root systems of structural inequality facing not only the survivor but community members (including the abuser), in addition to the injustices and trauma survivors undergo~\cite{Asad,coker,mia}. In our discussion, we will explicitly discuss how an SCTJ approach toward systemic change may be beneficial for designers.

\subsection{To Design \textit{Within} Existing Structures}
Alternative justice theorists and advocates have contrasting views on whether systems (\textit{e.g.,} law enforcement, legal systems) should be invoked to address injustice or avoided to dodge systemic harm \cite{coker, stubbs2002domestic}. Not only do TJ and RJ differ in their relationship with the CJS (as explained in \hyperlink{SCTJ}{Section 2.2}), but within the TJ movement advocates have called for abolishing existing systems and building new ones as opposed to reform, while others view the current time as an opportunity to accomplish change within existing systems \cite{kim2021transformative}. Within the context of DV, SCTJ expands and prioritizes the victim-survivor choices, including involving the State \cite{kim2021transformative,stubbs2002domestic}, because harm and abuse persist when DV is considered a private matter to be addressed within the family rather than a social issue that requires public and systemic intervention. 
Relying on community practices entirely away from criminal justice and public resources can contribute to further harm by allowing community members to enable abuse~\cite{coker}. In our data, victim-survivors' immediate communities stigmatized and shamed talking about abuse, denied DV's prevalence in the community and actions required on their leaders' and members' behalf, and pressured victim-survivors to tolerate the abuse rather than seek professional and legal help. According to advocates, survivors, and imams, two factors make it necessary to involve law enforcement: 1) local communities continue to enable abuse and 2) the shortage of religious institutions' services. 
A nuanced issue that Muslim victim-survivors commonly face is the discrepancy between religious and civil divorce~\cite{Macfarlane,RN195}. Because religious marriage contracts are not reflected in the civil system, a controlling tactic used by abusers is to divorce civically and maintain the religious marriage or vice versa. For example, when a woman is divorced legally but not religiously, her marital status upholds socially, depriving her of the opportunity to leave the abuser or find another partner. Whereas if their marriage is religiously bound away from the court system, fundamental rights such as alimony and child support are negotiable or denied as an extension to the abuse \cite{RN195}.
By not reflecting divorce in both systems, harm is inflicted where survivors are either kept hostage in their marriage or their legal divorce rights are ignored. Constructing DV as a public issue and crime stresses its seriousness, allows for communities to hold individuals accountable, and attends to the survivor's choice and need for external validation and intervention~\cite{coker,islam2018challenges}.

In our study, survivors frequently experienced denial and gaslighting, which is when a person is manipulated into not believing their own experience, by their communities when they spoke up against abuse.
Often, the abuser's financial and social prominence contributed to revictimizing the victim-survivor within her community and the justice system. For example, two imams stopped supporting P22 when she decided to press legal charges against her abuser; in another case, imams declined to support P06 because of her husband's religious influence within the mosque community. 
SCTJ acknowledges both the benefits and harms potentially inflicted by the system, however, leaving DV solely for the community to handle can bring more harm than good. We recognize the subtlety in balancing such claim; there were multiple ways where public services fell short of helping our survivors, including when shelters turned down survivors for health and capacity reasons; when pro bono lawyers refused cases lacking evidence of physical abuse (despite the prevalence of psychological, financial, and emotional abuse tactics that are difficult to document); and when the police discouraged victim-survivors from pressing charges knowing the abuser's financial affluence would deplete the victim's resources in court and cause her more harm. SCTJ attends to these consequences by focusing on strategies to change community norms and control systemic harm. In our sample, several cultural-based DV organizations contribute to preventative work aimed to change social norms and systemic practices through community outreach, imams training, legal advocacy, and mainstream service providers coaching.
Imams and advocates are taking matters into their own hands by speaking at sermons, community events, and outreach programs. However, the community needs to support interventions and heal from the internalized stigma and denial harming their communities. SCTJ provides grounds for community transformation, requiring contextualized, collaborative, and consistent long-term efforts on behalf of all stakeholders involved (\textit{i.e., }community members, abusers, faith institutions, non-profit organizations, mainstream service providers, and law enforcement). Abusers are to be held accountable and community leaders, law enforcement, and service providers must be trained to adequately serve DV victim-survivors according to their specific needs and circumstances. Aiming to achieve both; to design within existing structures while working towards building healthier and sustainable systems is possible. However, we work \textit{within} to preference the victim-survivor's choice and because current alternatives are less-developed and in exploratory stages \cite{kim2021transformative}.  
In our discussion, we reflect on designing for diverse experiences and backgrounds in vulnerable situations with the intention of not reproducing harm. We discuss designing for abuser accountability and behavioral change, and influencing community ideology as ways to transform the conditions enabling gender-based violence.

\subsection{To Center Survivor's Autonomy}\label{centersurvivors}
Often, in cases of social issues like domestic violence, victim-survivors are decentered, which we define as when a person's choices, preferences, and autonomy are disrespected and ignored. We can see this decentering happening in multiple ways, including in the abusive relationships and when seeking services, but can also happen in how designers approach design situations in terms of how to understand a problem or how to address a problem that the victim-survivor is facing. A way to correct for this decentering is
through "supporting survivors around their healing and/or safety and working with the person who has harmed to take accountability for the harm they have caused"~\cite{mia}. In the DV context, survivors are centered by individuals, communities, institutions, or policies when a survivor's autonomy is prioritized (\textit{i.e.,} their choice of when, what, and how services and support are provided). Where abusers take responsibility for their actions based on the survivors' needs, whether its community exclusion, financial responsibility, legal repercussions, or a sentimental apology with a commitment to behavioral change. In comparison to the criminal justice system where survivors are rarely centered, and restorative justice approaches where harm is restored by relying largely only on community practices, SCTJ encourages leveraging community, mainstream, and governmental support to sufficiently aid the person being harmed in regaining their lives away from harm. 

In our study, we found that survivors' needs were frequently deprioritized by the criminal justice system and their local communities. The criminal justice system is tied to DV, where once abuse is reported by the victim-survivor or through witnesses, it follows a strict line of procedures that do not necessarily align with the survivor's choices or priorities~\cite{harmreduction}. For example, when a victim-survivor seeks shelter services, criminal justice systems are often triggered, which may result in an abuser's deportation or detainment. This in turn deprives children of male father figures, leaves the survivor void of financial resources, and opens the survivor up for community shaming. On a logistical level, shelters were viewed negatively by Muslim survivors due to the fixed structures and rules shelters have. Our participants expressed victim-survivors' need for flexibility around meals (\textit{e.g.,} having halal meats, eating at sunset when fasting), space for daily prayers, and privacy away from men for women who wear the hijab \cite{sokoloff2005domestic}. Such needs are all rooted in their religious and cultural values, and accommodating their needs can minimize hesitancy and discomfort toward shelters. 
P11 stressed centering survivors by allowing them to be heard, understanding their traumas, and trusting in their decision-making. 
 SCTJ as an alternative model of justice provides space for supporting survivors with services based on their own definitions of justice while relying on non-punitive practices. Comparatively, advocating for a harm-reduction approach centers strategies taken by survivors to reduce adverse consequences, regardless of what the community or system might find appropriate \cite{koyama2001toward}.


Finally, centering domestic violence survivors in their help-seeking process can be done by providing female and religious support options within law enforcement and faith institutions. By female support options, we mean the possibility to speak to a female officer, imam, or service provider. Participants expressed anguish when speaking to male officials and imams about their abusive situation due to privacy concerns and the potential lack of understanding and support. 
Being sensitive to traumatized individuals' needs is crucial in their recovery process \cite{TIP}. When gender is a part of a traumatic experience, as in DV, victim-survivors can experience intense fear, anger, or discomfort in dealing with the same gender as the abuser~\cite{TIP}. Thus, part of centering victim-survivors is attending to their nuanced preferences during their help-seeking and healing journeys \cite{sokoloff2005domestic, chatterjee, goodmark2004law,oyewuwo2020black, oyewuwo2016american}. 

\section{Findings}\label{findings}
\begin{figure}[h]
 \includegraphics[width=\textwidth]{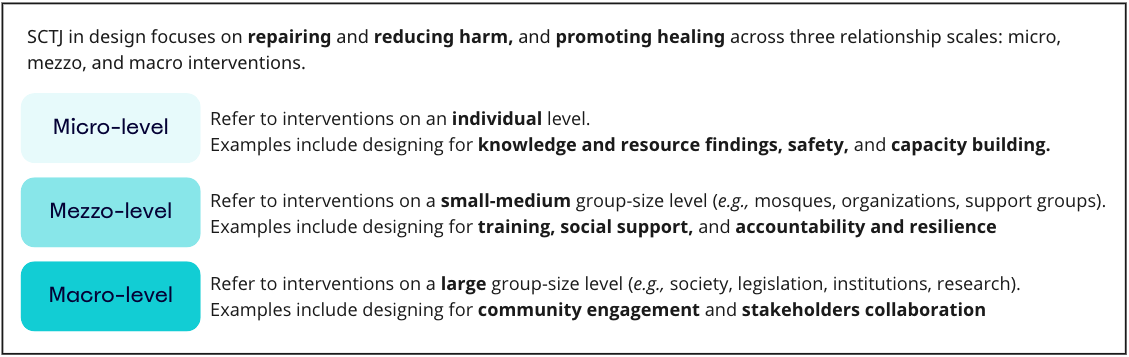}
\caption{Findings Summary: SCTJ in Design.}
\end{figure}
\hypertarget{findingsfig}{}

In this empirical section, we use insights from our participatory interviews to highlight possible intervention areas for reducing harm, repairing harm, and promoting healing within different scales of relationships for the DV survivors (a summary of the interventions is in \hyperlink{findingsfig}{Figure 1}). The second half of our interviews consisted of design activities focused on how the participants envisioned social computing technologies addressing specific challenges discussed in the empirical part of the interviews. We paid careful attention to any possible concerns or challenges anticipated from the prospective designs. 
Most of the technological transformations envisioned by our participants focused on repairing or restructuring relationships. By relationships in this context, we mean the roles and members involved in the victim-survivors' help-seeking and healing processes, which include the members who have done harm (\textit{e.g.,} spouse, in-laws), the person who's been harmed (\textit{i.e.,} victim-survivor), and individuals, institutions, and policies that influence the victim-survivor's experience.

Adapting from social work practice, we organized the suggested solutions based on their scale into micro, mezzo, and macro levels of interventions~\cite{bernstein1997social}. Micro-level interventions include individually dealing with the survivors. A mezzo-level includes interventions at the scale of smaller groups or organizations. In our context, mezzo-level groups could be mosques, anti-DV organizations, mainstream services, or support groups. Whereas macro-level interventions influence a larger portion of society, public policy, legislation, institutions, and research. By attending to the three levels of intervention, we adhere to the SCTJ values of changing the underlying conditions of abuse, holding community members accountable, and advocating for victim-survivors. 
Though social work practice has been criticized for normalizing power structures \cite{herz2012doing}, at its core, social work values integrity and challenging injustice, and is focused on individuals and society's well-being as a whole~\cite{nasw}. When examining social problems, we find social work practice and values to strongly align with HCI and view both fields as allies; social work in providing well-established practical knowledge to apply to social contexts and theories that can guide design, and HCI in supplementing interventions with an in-depth understanding and vision of technology's role and impact on interventions. 

\subsection{Micro-level Interventions}
By micro-level intervention, we mean interventions focused on the individual who has been harmed. Within our study, most proposed designs directly focused on victim-survivors' needs within three overarching categories: knowledge and resource finding, safety, and capacity building. In this section, we focus on how victim-survivors begin the critical work of repairing the relationship they have with themselves and the role that technology could play with repairing previous harm and offering up possibilities for new futures. In what follows, we will define and explore each category. 
  
\subsubsection{Knowledge and resource finding: }The top challenge reported by our survivors was the overwhelming process they undergo to find relatable and reliable information and resources online that suit their cultural views around domestic violence. Culturally sensitive content would expand DV's definition to include in-laws and family members~\cite{Dabby,kim}, rights and responsibilities from a religious perspective, and spiritual-based resources. Our participants reported using search engines to find religious information about their situations, only to find victim-blaming content. Searching for legal guidance presented them with an overwhelming amount of information that was difficult to digest, especially in stressful situations. 
To meet the victim-survivor where they are by minimizing the flux and irrelevance of information, P20 suggested a gradual influx where the tool is 
\textit{"geared towards Muslims particularly definition-wise… that you're informed with first, and then you have different options: (What do you want to do? Do you want to go to a therapist?). Go from less severe to more severe choices, the hotline first, and then all the way down to (here's your court option)."} 
To enable informed decision-making, victim-survivors need specific ways to receive and process information; where relevant information is presented step-by-step and freedom is afforded to choose what route fits them best.
 
To address content challenges and concerns, participants envisioned optimized search engines, dedicated websites, and mobile applications, that acted as a multi-lingual repository for DV resources. Where tailored information is presented based on one's demographic information (including religion, language, and location) and situational information (\textit{i.e.,} DV case, needs). Information needed included local supporting services and shelters aggregated textually or visually in the user's preferred language and tailored to their specific needs (\textit{e.g.,} spiritual-based counseling, interpretation services). 

Resources alone are insufficient for people going through abuse.   
Here, P07 envisions a socially constructed community for support and resources: 
\begin{quote}
"a community of people who are going through the same thing…where they are able to receive different resources, where they are able to meet one another or come together as a group, where they can brainstorm themselves way beyond what is needed or what is to do." P07, F, Case Manager. 
\end{quote}
Online social support groups are effective in providing emotional and informational support for people with shared experiences~\cite{chung2010benefits}. However, working in online social collaborative peer spaces opens up various, legitimate safety and privacy concerns, especially for victim-survivors [\textit{e.g.,} \cite{chen2022trauma, blackwell2017classification, schoenebeck2021drawing, Xiao2022}], we will discuss this more in depth in (\hyperlink{section.5}{Section 5.1}).


A trauma-informed knowledge-finding process entails validation, where victim-survivors' are listened to and believed, solidarity, where DV is acknowledged as a serious problem, peer support, where they are connected with fellow survivors~\cite{chen2022trauma}, and the victim-survivor's ability to preserve their autonomy. Here, SCTJ manifests in design by centering the user's autonomy in paving the way for victim-survivors to identify the abuse, make informed decisions, and feel supported throughout the process~\cite{chen2022trauma}.

\subsubsection{Safety: }The second category emphasized by our participants was the need to maintain survivors' safety. Safety here warrants being physically and digitally safe (\textit{i.e.,} away from technology-enabled abuse \cite{freed2018}) from the abuser and potential enablers of abuse, the ability to securely document, share, and report abuse, and to interact with empathetic and capable law enforcement and service providers.

Participants stated the need to alert victim-survivors of their abusers' physical proximity and if there is potential technology abuse, and learn how to maintain their safety online, which aligns with prior HCI work \cite{freed2018, freed2019}. An example is raised by P12: \textit{"If there's a way to set up some kind of an alarm to say (it looks like there's an added device on your system, or it looks like somebody is gathering your information). I think that could be really helpful"} P12, F, Program Manager. 
The need for platforms to notify users about potential risks has been proposed in prior work \cite{freed2019}, however, the issue of misunderstanding the information flow was found to be common for the average user \cite{freed2019}. Thus, visualization tools are suggested to help users build better mental models about the flow of information across their devices and accounts \cite{freed2019}. 

To design for safety is to balance visibility on ICT and online platforms and guide users on ways to minimize or remove digital threats. Contrarily, visibility, usability, and accessibility can become harmful when tools are obtained by the abuser, where they can use the same tools to stalk or further control the victim-survivors~\cite{freed2018, freed2019}:
\begin{quote}
"I would want to know if the abusive partner is going to be alerted. If I'm [the abuser] trying to have your messages forwarded to me and I try, but you have this kind of security system in place, am I going to be notified that there's a block? because if I'm notified, then I could question you, or I could harass you about it… and create a more unsafe, unstable environment" P12, F, Program Director. 
\end{quote}

Abusive behaviors are used by abusers to exert control over partners or family members \cite{mslmpowerctrl,dvPFP, rankine2017pacific},  thus, there lies a trade-off between publicizing DV-related tools and abusers potentially exploiting them for controlling tactics. HCI scholars suggested degrading usability for adversarial users (\textit{i.e., }abusers) after detecting unfamiliar abusive behavior or unfamiliar usage patterns of exploited accounts~\cite{freed2018}. To minimize abusers' exploitation of DV-related tools, participants mentioned disguising tools under utility or religious mobile applications to absolve any suspicion on behalf of the abuser.
P22 explains: 
\begin{quote}
    "you want the people who need it to find it, but then you need them to be safe! so what happens? when you install it [the app] changes names? … can I hide it as a Quran app and then when you open the screen it shows a surah [Quranic chapter], but if I click on surah three times then it opens the app?" P22, F, Survivor.  
\end{quote}
To avoid risking exposure, participants suggested designing discrete wearables (\textit{e.g.,} necklace, watch) (similar to ~\cite{brown}). Others wished for a texting feature rather than phone calls when reporting to or alerting authorities, which is currently not available in all States. For such interventions to work, access to reliable technology and knowing how to operate the tools are required, which may not apply to many victim-survivors. Lastly, integrating pre-agreed code words during online interactions as ways to uphold victim-survivors' safety was suggested, and is currently used by service providers to communicate with victims, especially during the times of COVID-19, when the abuser is often within close proximity.

In the context of DV, maintaining the user's privacy may clash with broader privacy and trust concerns for Muslims. For example, famous religious apps have been reported to sell users' information for state surveillance~\cite{muslimapp}. Indeed, imams, service providers, and survivors alike disclosed that fear of reporting abuse is a barrier to victim-survivors' safety within the Muslim community. By taking a survivor-centered transformative justice approach, ensuring privacy for marginalized groups involves considering the systemic concerns they have and how such concerns may affect their technology use and adaptability.

\subsubsection{Capacity building: } Lastly, we want to highlight the importance of building victim-survivors' capacity to regain their sense of self. By capacity, we mean educational, vocational, and financial knowledge and skills to sustain themselves away from the abuser. Many victim-survivors may have a reduced sense of their own agency and capacity because of the abusive situations. 

One way to begin building confidence is to seek counseling. Participants found it vital for counselors to be spiritually-aware: 
\begin{quote}
"I was very adamant about finding a Muslim therapist because I felt like a secular therapist can give me tools and they can give me strategies and they can teach me things of course, but I knew that I would need to mend other things that they may not necessarily relate to, or they don't understand" P19, F, Survivor.
\end{quote}
Other participants echoed P19's sentiment, where they found spiritual-based counselors' advice relatable to their experience and belief system, and answered questions from a faith and cultural viewpoint, for example "\textit{to explain to them from an Islamic standpoint, why is this happening? or where do I go from now? how do I deal with this stigma?}" P19. Whereas P14, a survivor, had an enriching experience with her Christian spiritual-based therapist and recalls an unanticipated moment of revelation when her therapist
mentioned \textit{"God doesn't give you things more than you can tolerate"}, which P14 immediately connected to a Quranic verse, helping her gain the clarity and strength at that moment to push through. 

Technology was seen as a tool to connect with others going through similar experiences, mentors, and therapists who are affordable and adequately understand the cultural nuances of certain communities. 
Social media helped P06 transcend physical limitations to escape and heal from an abusive partner: \begin{quote}
"I think women who are in DV get realization from other sisters, going to the village and building a village, we are going to our elders to pick advice for our personal life, I just made my village broader and bigger, and the way I'm doing that is through networking on Facebook" P06, F, Survivor.
\end{quote}
Social media groups were used by victim-survivors to feel less isolated, build their understanding around their situations, and offer advice and help to other victim-survivors in return. 

Building financial literacy and providing language access for victim-survivors through cost-effective technologies was emphasized by social workers. Currently, such technologies are offered by big tech companies, however, are unaffordable for the people who need them the most: \begin{quote}"language access is available in a lot of spaces, it still didn't trickle down to the basic consumer. AT\&T has I don't know how many languages, it used to be \$4/minute…the vulnerable people need it the most, and I think it's an evolution" P04, F, Executive Director. \end{quote}Lack of access to capacity-building services and tools is a significant barrier to fleeing abuse. We join our participants' plea for tech companies to make such services affordable and accessible for socially and financially vulnerable populations. Affording literary and capacity-building services contributes to a SCTJ approach by enabling behavioral change on behalf of the victim-survivors, where they rely less on the abuser, and alleviate harm.

\subsection{Mezzo-level Interventions}

Mezzo-level interventions include intermediate-scale interventions within small groups (\textit{e.g.,} family, community, neighborhood), and by establishing credibility and accountability within organizations through building professional capacity and evaluation. From an HCI and DV perspective, mezzo-level thinking may help designers focus on encouraging social change at the group level. In this section, we design for group-level change by focusing on key stakeholders who likely have larger effects on the group. Mezzo-level interventions are critical to promoting community healing, with the underlying assumption that people who know better, can do better \cite{interventions2012creative}. In what follows, we will highlight these stakeholders and the possibilities and challenges of design for training, social support, accountability, and resilience.

All three participant groups (\textit{i.e.,} service providers, survivors, and community leaders) stressed the need for standardized DV training of key community members and leaders who are at the forefront of responding to DV cases. Within religious institutions, mosque structures commonly have a female liaison (often the imam's wife), who is easily accessible to female attendees, assists with their needs, and acts as an intermediary between community members and the mosque leader or board members. Female liaisons are often present and involved during client sessions with the imam. The liaison's role is significant, though their constant presence with the imam and his female guests is religiously-rooted where being privately alone with a member from the opposite sex is prohibited, participants expressed the need to speak to a woman about marital issues. For imams and their liaisons to qualify as a central point of contact when encountering DV victim-survivors, training must involve specialized trauma-informed spiritual counseling and knowledge-building around available local and national resources and their different roles (\textit{e.g.,} hotlines, credentialed therapists, shelters). 
 
In our interviews, participants often mentioned a need for anti-bias training for community leaders and members alike as bias poses a significant barrier to adequate service provision. P08 mentioned a common type of harm found through service provision she calls "the auntie syndrome," where
\begin{quote}
    "people having improper training, they'd jump in and tell the person what to do, and that's just moving the power and control to us [service providers]! We are taking from whoever is doing the harm and making it ours! Assuming that we know what's best and that this person doesn't have agency and cannot make decisions for themselves" P08, F, Advocate
\end{quote}
While the "auntie syndrome" may come from a place of good intentions, ultimately, it is an unhelpful behavior as it does not center the victim-survivor's wants and needs nor builds up their capacity to act for themselves. Advocates and service providers oftentimes come from backgrounds of survivorship themselves~\cite{hammer2019peaceful}, are exposed to traumatic cases, and experience high workloads, leading to secondary traumatic stress and burnout, which may compromise their ability to serve clients effectively \cite{kulkarni2013exploring}. Working within DV services is a strenuous process, requiring patience, expertise, and self-reflection. Thus, for a survivor-centered approach, bias and trauma-informed training are essential for service providers to attain their full potential when serving DV victim-survivors. 

Training direct and indirect stakeholders matters because it may reduce the risk of re-victimizing survivors from shaming and blaming and giving insufficient or inappropriate advice (\textit{e.g., }to stay and be patient). However, an underlying concern was that service providers might lack empathy, as P07 puts it:
\begin{quote}{"but that [education] only does so much, to be honest. You need to make sure those people who are receiving that training actually have the empathy to care about their clients who are depending on them."} P07, F, Case Manager.
\end{quote}Empathy here means the openness and willingness to develop a shared understanding of their client's experiences, and patience for the mistrust that may exist within communities due to historical and institutional factors~\cite{gottlieb2021case}. Further, within organizations, shelters, and government facilities, problems of internalized patriarchy and white supremacy~\cite{mia, sokoloff2005domestic}, and lack of cultural empathy among mainstream service providers were prevalent. Within the context of DV, cultural empathy refers to how service providers can commit to an ongoing process of self-awareness, view culture as victim-survivors may see it rather than only from their own constructs, and continually consider the social systems that shape the reality for victim-survivors~\cite{gottlieb2021case}. 

For technology design to improve training experiences, participants desired to transform virtual platforms to accommodate safety and synergy through design. By encouraging rapport-building and safe sharing around trauma within virtual training and support spaces, technology can serve as an emulating space for in-person interactions.
For mainstream service providers, participants suggested providing cultural-related DV content in the form of learning modules using games, scenario-based storytelling, and visual simulations to account for the broader DV dynamics where users: \textit{"take an in-depth look at different situations and not just the stereotypical physical violence, but maybe offering simulations, so law enforcement can be trained in different DV situations"} P12, F, Program Manager. 
While technology alone will not eradicate internalized biases nor replace needed on-the-ground outreach efforts, technology's role may be to supplement and support efforts where empathy exists, and policies are enforced to encourage anti-bias growth and cultural empathy. 

To transform relationships on a mezzo level, participants called for using technology to encourage community accountability and foster interconnection. One way to encourage community accountability is by embedding feedback forums for victim-survivors to review and evaluate the services received from organizations and mainstream service providers, providing a way for survivors to voice their concerns and increase the visibility of service quality~\cite{RN31}. 
Sustaining community giving through crowdsourcing is a way to use technology to encourage community commitment to bettering victim-survivors conditions. Participants P02, P03, and P04 proposed using crowdsourcing in sponsorship-based platforms, where sponsors donate temporarily or over an extended period to cover survivors' basic needs (\textit{e.g.,} single night or monthly housing, transportation, groceries, clothing, courses) for a low-cost subscription fee for victim-survivors. P03 emphasized that professional moderation and confidentiality are required for such service: \textit{ "We [service providers] would be in charge of it of course"} and \textit{"[housing] stays anonymous, they don't know where the building and locations are"} P03, F, Counselor. The donor-matching model was intended to not only support victim-survivors financially but also to help overcome mainstream administrative barriers such as the need for documentation and proof of address for victim-survivors to qualify for financial assistance. 

Lastly, to foster accountability and resilience for both the victim-survivor and abuser, the concept of community \textit{pods }was proposed. A \textit{pod} involves relationships "between people who would turn to each other for support around violent, harmful and abusive experiences, whether as survivors, bystanders or people who have harmed. These would be the people in our lives that we would call on to support us with things such as our immediate and on-going safety, accountability and transformation of behaviors, or individual and collective healing and resiliency"\cite{pods}. Participant five elaborated on this thought by sketching a tool where 1) survivors can track their case progress, name the harm they experienced, involve community members, and help manage resources to support their well-being and healing, and 2) community members learn how to support and hold accountable both the survivor and abuser in dealing with abuse (\hyperlink{fig2}{Figure 2}). 

\begin{figure}[h]
\hypertarget{fig2}{}
\centering
\includegraphics[scale=.60]{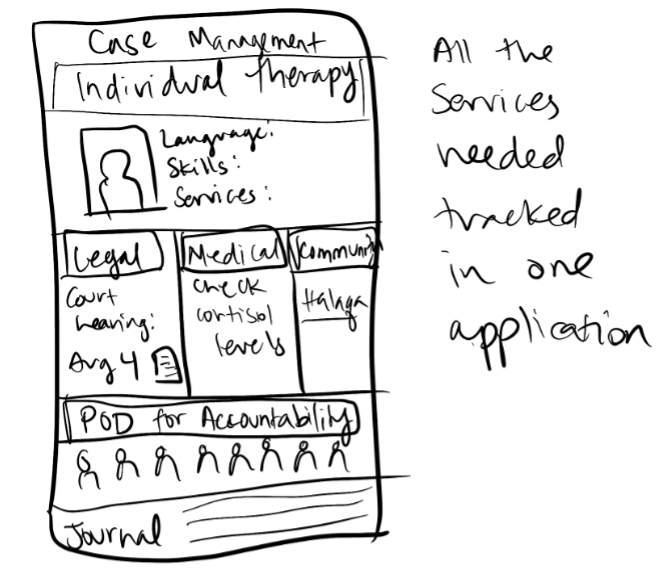}
\caption{Participant Five's Sketch of Survivor's Personal Management Tool.}

\end{figure}

Participants expressed serious concerns around their suggested interventions, including computer literacy, access to technology, and the learning curve of adapting new tools. Privacy and safety are critical in crowdsourcing and sponsorship models, where exposure to harm can occur through abuser stalking or exploitation of contractors who provide their services to victim-survivors (\textit{e.g.,} landlords). Designing for mezzo-level interventions requires profound expertise to understand and address triggers and biases\footnote{Several organizations have been working on training materials (\textit{e.g.,} \href{https://www.peacefulfamilies.org/}{the Peaceful Family Project resources}) that can be redesigned and disseminated.} and careful consideration of how to portray victim-survivors without reinforcing stereotypes. Additionally, such interventions can be overly ambitious considering the reality of the human service sector, where resources are depleted and employees are overworked \cite{kulkarni2013exploring}.

\subsection{Macro-level Interventions}
Macro-level interventions influence a larger portion of society, public policy, legislation, and institutions. In our findings, macro interventions focused on involving the broader community and promoting collaboration between key stakeholders (\textit{i.e.,} Muslim service providers, faith institutions, and mainstream service providers).

Social workers emphasized the need for DV advocates and faith leaders to proactively encourage and prepare the larger community to understand DV signs, learn about available social services, and how to offer help within their immediate and broader circles. P21, a survivor and DV advocate, suggested a "train the trainers" model\footnote{P21 was trained through an advocacy program offered by the city of Dayton, OH, where community members are trained to offer resources and a safe place for victim-survivors.}, where appointees across the community are equipped with ways to direct victim-survivors towards adequate local resources, which decentralizes imams as the primary points of contact. Participants suggested reaching members of the community through a dedicated WhatsApp hotline or education modules integrated into social media platforms. 

To overcome technology access barriers, internet hotspots and portable devices for houseless populations across cities were suggested: \textit{"it's hard to keep track and stay connected to certain houseless individuals because they are constantly moving or changing location, so something that doesn't require that you constantly have something on them but it's an instant type of connection"} P05, F, Program Manager. Mobilizing technology access is crucial in cases where victim-survivors are deprived of access by their aggressor or lost their homes as a result of the abuse. 

On a broader level, faith leaders and social workers called for "building bridges," where a collaborative relationship exists between members of faith institutions, Muslim service providers, and mainstream service providers: 
\begin{quote}
"the perfect relationship would be if we have the religious expertise and leadership coupled with the therapeutic expertise and tools in that dynamic and they work in conjunction with each other"
P19, F, Mosque Liaison.
\end{quote}
Our interviewees highlighted the responsibility faith institutions must hold in actively engaging with DV coalitions and service providers to build faith leaders' understanding of DV and services layout, and establish a working relationship where open communication exists between all key stakeholders to provide joint and holistic interventions for survivors~\cite{Xiao2022}.

Building bridges among various stakeholders not only helps in better serving victim-survivors but also in empathy-building and minimizing biases, as P10 passionately narrates:
\begin{quote}
"…sometimes it's just being with people, it's like "oh they're not that different!" or "I know someone who's Muslim, or I know someone who's Jewish"…that's what builds the sense of engagement, builds the sense of empathy, and then folks are not so strange…when we care it's not that hard." P10, F, Social Worker.
\end{quote}
Cross-cultural interactions build a shared cultural understanding, where differences are minimized and commonalities are amplified across groups, nurturing a culture of care that is greatly needed in serving at-risk communities. 

The concerns participants had around macro interventions were mainly around policy obstacles (\textit{e.g.,} the difficulty to integrate law enforcement into collaborations with nonprofits and mosques). Also, the fear of reinforcing systemic injustice through technology (\textit{e.g.,} tech surveillance ~\cite{muslimapp}). Lastly, the difficulty of incentivizing service providers who are consumed with a heavy workload or community leaders who do not see DV as a priority to change their practices.

Addressing challenges on a macro-level requires long-term, localized, and ongoing advocacy. Thus, we see technology as a supportive tool in interventions~\cite{brown} and outreach, not as a replacement for on-the-ground advocacy work. By designing on a macro-level scale, we align with TJ value of community transformation, where underlying conditions of oppression and harm are addressed through social and systemic change.   

\section{Discussion}\label{discussion}
Research in HCI has built on multiple facets of DV-related interventions; tech abuse~\cite{freed2017digital, freed2018, freed2019}, survivor support~\cite{RN8, bellini2019mapping, rab21}, perpetrators' behavioral change~\cite{mechanisms, Bellini2020}, and trauma-informed computing~\cite{chen2022trauma}. In this work, we dive further into how to center survivors facing compounded identity-based oppression and trauma while respecting their societal and religious preferences. 
We started this project focusing on the challenges Muslim victim-survivors face in their formal help-seeking process when interacting with mainstream services, law enforcement, and the legal system in the US. However, speaking to service providers, community leaders, and victim-survivors revealed that the problems abound when trying to seek help; in our data, we found that victim-survivors often undergo revictimization through mainstream service providers, community members, and religious authorities during reporting, help-seeking, and healing processes. 
To center survivors, we need to move away from the individualistic intervention model to a collective one; where intergenerational trauma, historical and current rampant discrimination, sociocultural factors, and consequences of political conflict (\textit{e.g., }immigration) are all taken into consideration for victim-survivors, abusers, and community members. By taking an SCTJ approach, we work toward a collective intervention model that minimizes the burden often put on victim-survivors and provides them with immediate and long-term support~\cite{genfive, harmreduction}. \newline
Our conceptual \hyperlink{TJconditions}{(Section 4)} and design \hyperlink{findings}{(Section 5)} findings align with multiple works in the HCI literature, including social justice-oriented and prefigurative design~\cite{SJOR, Asad}, the six principles of the trauma-informed computing framework \cite{chen2022trauma}, the four mechanisms of moral responsibility to challenge harmful behaviors \cite{mechanisms}, and the Islamic feminist stance to inform designing for agency practices within oppressive conditions \cite{rab21}. In this section, we build on HCI work and use a SCTJ lens in correspondence to the macro, mezzo, and micro strategies (in \hyperlink{Section 5}{Section 5}) to advocate for a survivor-centered approach by: 1) designing for harm reduction, 2) designing for accountability, and 3) designing for systemic change, to supplement victim-survivors' help-seeking and healing experiences through design. 

\subsection{Designing for Harm Reduction}\label{harmreduction}
In this subsection, we focus and expand the conversation around \textit{meeting people where they are} in their journey for change \cite{mechanisms} and recommend designing for the victim-survivor's informed choices \cite{chen2022trauma} by considering harm-reduction as our guiding light, despite designers' (or other stakeholders') potential conflict with the victim-survivor's decision~\cite{harmreduction}.
Harm reduction within DV is a non-linear, self-identified, and internally motivated process unique to the victims-survivor to help "reduce, avoid or escape violence and to minimize its effects"~\cite{douglas}, and involves developing strategies that attend to "individuals' and communities' goals, needs, strengths, and deficits" ~\cite{harmreduction}. Here, we address the conflict in balancing how we, as researchers and designers, can be sensitive to the needs and decisions of our users while considering the consequences of trauma and abuse such as the emotional and physical deleterious effects of DV, dissociation, and decision-making paralysis, as reported by scholars \cite{van2022posttraumatic} and our participants. In other words, in situations where one's agency is constantly dispossessed, making critical life decisions can be overwhelming, burdensome, and result in serious harm (\textit{e.g., }staying and tolerating the abuse). However, when professionals make decisions for victim-survivors and not \textit{alongside} them, such professionals are replicating oppression and causing further harm (\textit{e.g., }separating from the abuser can result in further violence) ~\cite{harmreduction}, which we do not wish to replicate in the design process.  \newline
To design for harm reduction, we must first develop a deep understanding of our participants' needs, choices, and community practices related to DV~\cite{harmreduction}. To understand the unique needs of Muslim victim-survivors, in both the problem formulation and analysis phases, we develop empathy toward the survivors' choices and needs by following an Islamic feminist sensitivity toward understanding the historical, cultural, and political contexts of Muslims in the US \cite{rab21}. Due to historical discrimination and in response to political events \cite{ali2017impact}, the mosque's role in the US acts as a spiritual sanctuary and a community-building and service provision space~\cite{hammer2019peaceful}. Our participants refrained from resorting to or navigating governmental services, preferring to tolerate the abuse or reach out to community leaders (imams) when consequences get dire. In fact, imams may have a tight coupling with the local police stations to provide contextual support for DV cases. Another aspect of community practices was resorting to male allies to confront abusers or advocate for the victim-survivor \cite{rab21}. In some cases, male allies were acquaintances of the abuser, who leveraged the trust and social standing they have with the abuser, while in other cases, they were community members who voluntarily confronted abusers at their homes to stop the abuse.
\newline By taking Islamic feminist and harm reduction stances, we support and consider designing around women's agency in seeking cultural, religious, and spiritual guidance to overcome, deal with, and heal from abusive situations.
For informed decision-making, we can start by advancing the current DV tools, such as the national DV hotline website\footnote{National Domestic Violence Hotline. The Hotline. (2022, June 13). Retrieved November 21, 2022, from https://www.thehotline.org/}. While it provides context-based information and a local-resources search option based on services and population (including "Muslim"), there remains space for improvement. For inclusivity, a way to start is to include content translations of both widespread and niche immigrant spoken languages (currently, it only provides English and Spanish). Another is to expand the definitional information to account for broader forms of abuse (\textit{e.g.,} spiritual abuse), cultural-based information countering abuse (\textit{e.g.,} rooted in religious beliefs), and enablers of abuse (\textit{e.g.,} in-laws, community leaders).  
Knowing imams are first responders \cite{Karamah} and the role and involvement of community members, we recommend designing for imams and male allies~\cite{rab21}. One imam we spoke to followed a transformative mediation approach where she believed victim-survivors had the power within them to reach the decisions that worked for them, and her role was to help them tap into that power instead of offering them solutions.
Another imam prioritized empowering women in the community by ensuring their safety at home, providing positions within the mosque communities, and connecting them with the resources they needed. While another worked directly with police officers, issued restraining orders, and followed up periodically to monitor the situation and reassure their support. Imams offered mediation with abusers and spiritual support for victim-survivors. Thus, including community resources such as spiritual leaders and social networks, on the hotline webpage, as an intermediary step to seeking advice may help make the process less intimidating for individuals who are in non-physically threatening situations or their sensemaking process \cite{Xiao2022}. Nevertheless, the imams' roles are limited at best and often cause further harm. Thus, we view extending designing for responsibility~\cite{mechanisms} as a harm reduction strategy vital in this space, where non-abusive behavior for community leaders and members is encouraged. One way to design for responsible faith leaders is by creating a space to build their knowledge on how to support victim-survivors adequately and directly access credible sources, professionals, and care providers (\textit{e.g., }a centralized portal with evaluated resources). Another way is to build fact-check simulators to enhance and test their knowledge, training them to provide sufficient advice and practice their actions in a safe environment where potential consequences are presented based on trauma-informed and spiritually rooted information. For community members, reducing harm can be by providing spaces for abusers to interact with and be mentored by vetted imams and influential community members, and for imams to be mentored by trained and senior imams. Lastly, we suggest expanding the concept of providing peer support~\cite{mechanisms} to include diverse community members' roles who are part of the DV ecosystem; for example, having an open line of communication between social service providers, imams, police officers, and advocates, sharing resources and best practices, and collaborating in interventions when needed with checks and balances to limit power differentials. \newline
In designing for harm reduction, we encourage designers to aim toward enhancing safety and reducing negative repercussions according to survivors' choices, whether by relying on religious support, community practices, or governmental services to alleviate the ramifications of abuse. It is difficult to guarantee a harmless route when dealing with DV; however, centering survivors' agency and working on their capacity to make informed decisions is key in providing DV justice.

 \subsection{Designing for Accountability}

A distinctive attribute of SCTJ is that the person who has done harm (\textit{i.e.,} abuser) is perceived as a product of societal or systemic harm, meaning they too have been subjected to harm by their communities. The implication is that because their behaviors are learned, they can be unlearned \cite{mia, chatterjee}. For example, gender-based violence is viewed as an outcome of patriarchal beliefs especially around masculinity \cite{api}, harming both males and females. Masculinity, which "refers to the roles, behaviours and attributes that are associated with maleness and considered appropriate for men"\cite{UNwomen}, is often associated with being tough and disassociated from emotions, resulting in emotional immaturity and communication problems. Systemic discrimination is another form of harm abusers undergo in the context of DV in the Muslim community (see \hyperlink{section.2.1}{Section 2.1}), which may result in unresolved traumas leading to abusive behaviors \cite{maldonado2021does}. SCTJ prioritizes the victim-survivor's wellbeing over the abuser's rehabilitation, and highlighting the abuser's circumstances is not a justification of abuse. Instead, such a perspective provides space for restoration through accountability and long-term support, and grounds for changing the conditions that engender and sustain the abuse. 
Accountability is an SCTJ value perceived as a critical mechanism of justice and a powerful tool of transformation~\cite{genfive}. Within relationships, accountability is defined as "willing to interrupt problematic behaviors or dynamics and then support a process for transforming those behaviors"\cite{genfive}, and requires: 1) stopping immediate abuse, 2) acknowledging the harm done and its negative impacts on individuals and the community, 3) making appropriate reparations, and 4) committing to a non-abusive future. In the context of DV, accountability involves the enablers of harmful behavior, including the abuser, community members and leaders, and service providers. In transformative and restorative justice models, the abuser's accountability process may involve other community members as part of the support system. However, the community is only involved when community members' capacities are built to "support the intervention, as well as heal and/or take accountability for any harm they were complicit in." Communities can be complicit in DV by "ignoring, minimizing or even encouraging violence"\cite{interventions2012creative}. 
Thus, accountability in SCTJ is a dual process; abuser accountability and community accountability, as a route to achieving collective liberation and individual justice~\cite{genfive}. The concept of community accountability is also rooted in Islamic thought (\textit{i.e.,} the notion of \textit{ummah}), where Muslims are encouraged to take care of community members to the best of their ability~\cite{ummah}.
Here, we see a cultural alignment between Islam and the \textit{pods} concept, which are social structures that provide immediate and on-going safety, accountability, healing, and resilience to both parties (\textit{i.e.,} the survivor and abuser) \cite{pods}. The significance of pods within SCTJ is to hold both parties (the survivor and abuser) accountable, aid in their healing process, and ultimately transform the surrounding community to promote anti-DV beliefs and foster adequate healing resources. The pods concept can be integrated within social networking groups, for example, by including resources (\textit{e.g.,} vetted therapists, advocates) and information around healthy masculinity and families, peaceful relationships, or emotional coping in groups related to moral or social support (\textit{e.g.,} social initiatives and mosque pages).

Within HCI, accountability has been discussed as a dimension of social justice-oriented research~\cite{SJOR}, defined as holding people who encourage or benefit from oppression responsible and setting restrictions and appropriate penalties~\cite{SJOR}. In \cite{mechanisms}, Bellini et al. follow the notion that abusers can unlearn abusive behaviors, and provide moral responsibilities for perpetrators to refrain from abusive acts. Specifically, designing for the moral mechanisms of \textit{self-awareness}, where abusers acknowledge their role in causing and accepting blame for their abusive actions, and \textit{acknowledging the extent of harms} of their behaviors on others, both support the second step of the SCTJ accountability process (\textit{i.e.,} acknowledging the harm done and its negative impacts on individuals and the community). For example, by showing examples of children or older family members (\textit{e.g.,} parents or in-laws) addressing the consequences harmful behavior had on them and relating it to the actions of the abuser, the extent of harm is depersonalized and presented away from direct blame.
When designing for self-awareness of abusers, we propose designing for ways to acknowledge not only the harm they have done but how they have benefited and were harmed by the community and the system that may have contributed to the abuse. For example, by connecting them to reformed abusers and professionals who can walk them through alternative futures.
A central aspect of accountability is properly apologizing by admitting the harm inflicted and providing reparations to compensate for harm and as a form of commitment to change, especially for those who wish or have no other choice but to remain in the relationship \cite{accountability}. Technology can be leveraged to guide the apology process through textually analyzing written forms of apologies and providing digitally generated guidance or connecting abusers to mentors who can help them in forming their apologies and reparations.

When it comes to community accountability within marginalized communities, it is not a straightforward process. In the literature and our data, we see the complexity of what historical discrimination can lead to; our participants were receiving strong messages from their communities around the need to conceal conflicts from the public and avoid interacting with the criminal justice system in fear of reinforcing stereotypes about Muslims and Islam, and of triggering further harm through the system (\textit{e.g.,} unlawful charges, deportation). Another issue is confronting deeply rooted patriarchal beliefs, such as justifying aggression as a form of discipline or viewing divorce shamefully as an act of destroying the family unit. Intergenerational trauma and patriarchal ideology lead to cycles of abuse; where victim-survivors fear reporting abuse and are often faced by denial within the community when they do report, making it hard to hold members accountable. Thus, in order to build the community's capacity, extensive work needs to be done to heal intergenerational cycles of trauma and abuse. Advocates and specialized organizations are doing this work on a small scale, training their staff to identify their traumas and learn how to dissociate them when engaging with victim-survivors. However, these are dispersed efforts that do not encompass the mass need. Such work requires change on policy and grassroots levels. Organizations, faith institutions, and mainstream services handling DV require
culturally-sensitive trauma-informed training for their employees, board members, and volunteers, and policies that enforce and evaluate such activity. A way to build community accountability is to focus on advancing technologies to bring people into an environment sensitive to their needs and triggers and focused on trust-building. Another aspect is to think of ways to incentivize community members (\textit{e.g.,} imams, service providers), whose capacities are overwhelmed,
through design. Though this is not in the scope of this paper, we find incentivization to be key in designing for community accountability.

Lastly, accountability in design starts with the researcher. Researchers carry the responsibility of accurately representing the participants' sentiments, by co-collaborating throughout the design process. Scholars have emphasized the need for streamlining academic research, where research findings are disseminated through social media, blog posts, and traditional media for the non-specialized recipient to benefit and integrate findings into their personal and professional practice \cite{fiesler}. Further, as scholars, it is our responsibility to work on "de-othering" marginalized populations; through showing the historical conditions they face, presenting cultural nuances away from reinforcing stereotypes, and working alongside them towards their goals and aspirations. 




 \subsection{Designing for Systemic Change}
Anti-DV organizations' interventions are ongoing, lengthy, challenging, and aim toward long-term results~\cite{hammer2019peaceful}. In our data, we see examples of this, as a well-established organization is now yielding the fruits of 30 years of prevention initiatives with local police. Slow, evolving, and cumulative work conflicts with the fail-fast corporate logic and design "solutions." Where technology is developed and is perceived as a means to solve problems almost instantly, which is mostly not feasible in social problems \cite{wickedkolko}, and worse, may cause more harm than good \cite{RN68}.  
Numerous HCI scholars have discussed sustainable social change and design frameworks~\cite{SJOR, Asad,schoenebeck2021drawing, costanza2018design}.
We anticipate and commit to conflict where we are upfront about our goals and recognize our limitations within the process \cite{SJOR,Asad}. Following such sensibility in our work, though we encouraged our participants to think of "magical" solutions, we articulated our goals as shifting the conversation within the technology design community and providing guidelines for future developers, ultimately, narrowing the gap between current tools and Muslim victim-survivors' needs. Also, we acknowledged with our participants that there is not one "right" answer, instead, there are many pathways for change \cite{SJOR}. 
Aligning with the social justice-oriented interaction design and prefiguative design approaches to design for systemic change~\cite{SJOR, Asad}, we partnered with different stakeholder groups who have "experiences with and/or work to end oppression"\cite{SJOR} in formulating our problem space (\textit{i.e.,} the challenges victim-survivors face) and in producing design directions. Our work moves away from product-based concerns to focusing on collective practices and sociopolitical concerns and commits to designing for transformation, recognition, and "healing via transformative justice," where we highlight unjust practices and policies, value designing for evolving relations that produce inequalities \cite{SJOR}, and enable community partners to envision "what healing looks like for them"\cite{Asad}. In our PD sessions, building bridges between various stakeholders was proposed to establish a working relationship and open communication between key stakeholders that would adequately serve victim-survivors instantaneously while they are reaching out for assistance. This is a valuable space to explore in survivor-centered design. However, factors to consider in the design process include ways to collaborate in building and maintaining a database of resources, contacts, and unified definitions and values for intervention, building visibility and transparency to show the user who is available and willing to assist, and most importantly, vetting and training members who are part of this platform to ensure they are all working towards a healthy intervention. 

During the PD part of the interviews, we were faced with a conflict; on the one hand, the scale of the problem was massive compared to the technical possibilities both the participants and researchers were trying to envision. For example, two of the convoluted consequences of abuse found were undocumentation and transnational abandonment (see \hyperlink{appendix}{Appendix} for definitions and examples). Further, survivors came from diverse religious, cultural, and racial backgrounds, contributing to different experiences within both their communities and the system. In our sample alone, we had nuanced cases of religious and ethnic identities (\textit{e.g., }converts, Muslim-born, immigrants), each experiencing different biases and benefits within their immediate and broader communities.  P06, a white convert woman, was well aware of the privilege she had over immigrant women of color with low English proficiency when interacting with mainstream service providers, law enforcement, and the judicial system. Whereas P19, a Muslim-born African American, initially refused to reach out to the police because \textit{"in African American culture you don't call the police unless you have to call the police!"}, causing her to regret not creating a record earlier to limit further abuse. 
We see committing to conflict being beneficial not only in the researchers-participants and among stakeholders' relations \cite{SJOR}, but also within oneself as a scholar or researcher. Where we learn to balance humbling ourselves without trivializing the problem; by 1) acknowledging the massive reiterative efforts and expertise needed to work with intersectional experiences, 2) collaboratively translating what is feasible into design while taking into account the potential damage of inadequate design, and 3) admitting and committing to the realization that we cannot tackle such massive social problems, but rather appreciate the small steps we are paving within the cumulative evolving process of social justice. We would like to use this space to invite the community to discuss: How can we make long-term sustainable work more rewarding in research and design? We suggest incorporating mandatory ethics classes adjacent to coding courses, standalone ethics boards in big tech, and encouraging nuanced research contributions around marginalized communities and underrepresented populations through funding and inclusion into top-tier HCI and computer science venues.
Creating systemic change requires policy-level research and ethics lobbying, which starts in our academic institutions and industrial corporations. Technology is no longer a standalone tool; it is shaping our political views and realities~\cite{fujiwara2021effect, chou2017influence}. Thus, serious interdisciplinary work is needed more than ever. 

Lastly, the purpose of SCTJ's focus on the underlying conditions enabling the abuse is to remain critical of harmful structural and social aspects and, ultimately, change them. Within the early TJ movements (\textit{e.g.,} INCITE!, generationFIVE) came a strong opposition towards state violence and a commitment to abolishing the criminal justice system. Slowly influencing national DV and sexual assault coalitions committed to strengthening police intervention policies to reconsider their orientation towards state involvement. In practice, applying theories of justice to nuanced real-life situations looks different in each case \cite{coker} and requires expert judgment. Existing programs that use a TJ process do not reject state power, but primarily rely on supporting survivors, altering abusers' networks, and "processes that link gender ideology and subordination with experiences
of racial subordination and colonization" \cite{coker}. Victim-survivors choose and benefit from state services regardless of the potential racism and Islamophobic attitudes they may endure \cite{stubbs2002domestic}, and service providers, allies, and advocates tailor their responses based on the dynamics of each case. Further, TJ experts find these unprecedented times of protests against police brutality and political interest in change as an opportunity to "rapidly build the infrastructure necessary to nurture a robust transformative justice future" \cite{kim2021transformative}. Thus, it is necessary to view alternative justice theories as ways to expand our options to center and serve survivors rather than limit them (\textit{e.g.,} by only relying on community praxis), and use such theories to build sensitivities towards individual, social, and systemic justice within research and society \cite{kim2021transformative, coker}.

\section{Conclusion}\label{conclusion}
In this study, we investigated Muslim victim-survivors' challenges in their help-seeking and healing processes in the United States and co-collaborated with our participants to exhibit the possible roles technology can take in minimizing the obstacles they face. We found that victim-survivors are often revictimized by their immediate communities and the criminal justice system. 
We empirically showcase when and why using a survivor-centered transformative justice (SCTJ) approach may be productive as a theoretical and design lens.
We demonstrate how SCTJ manifests in design
by targeting three group-level scales of intervention: micro, mezzo, and macro-level design interventions. 
Based on our findings, we discuss ways to center victim-survivors through supporting their harm reduction strategies, equipping them to make informed decisions, holding abusers and community members accountable, and laying the grounds for social and systemic change to undo behaviors and conditions upholding DV in marginalized communities.

Supporting the parties involved in DV dynamics while protecting victim-survivors can be challenging and may not align with universal human rights or designers' values. However, we argue that navigating the social, legal, and structural landmines with least harm is through centering survivors' autonomy and supporting their decision-making process. 
\section{appendix}\hypertarget{appendix}{}

 \begin{itemize}
   
  \item \textbf{Undocumentation:} 
An undocumented individual is "anyone residing in any given country without legal documentation. It includes people who entered the US without inspection and proper permission from the government, and those who entered with a legal visa that is no longer valid"\cite{undoc}.
In the context of DV, Survivors can face being undocumented by marrying into an undocumented spouse, through the abuser controlling access to legal documents, or as a byproduct of a cultural and systemic clash; where a married man marries another wife and brings her to the US under a temporary visa, but because polygamy is banned by state law, the second wife becomes undocumented when the visa expires. Undocumentation leads to a series of consequences, including ineligibility for governmental aid and services, inability to get alimony or become financially stable, and homelessness, which can impede the process of gaining financial benefits to support themselves and their families upon separation.

  \item \textbf{Transnational abandonment }
is "a form of domestic abuse when vulnerable immigrant women are abandoned in their country of origin by their husbands"\footnote{Kymal , M., \& Nagarajan-Butaney, A. (2021, November 18). How Priya won a second chance at her American dream: A story of transnational abandonment. India Currents. Retrieved September 8, 2022, from https://indiacurrents.com/a-story-of-transnational-abandonment-how-priya-won-a-second-chance-at-her-american-dream/}.
 If they have children and by the time the abandoned finds a way to get back to the US, most likely the court will not grant the mother primary child custody under the conviction of child negligence (if more than seven days). As a result, the victim-survivor's burdened with obligatory child support, when typically, the survivor is financially unstable \cite{RN195}.
 
  \item Other related trends in abuse are explained in \cite[p.23--26]{RN195}.
\end{itemize}

\bibliographystyle{ACM-Reference-Format}
\bibliography{references}


\begin{thebibliography}{118}


\ifx \showCODEN    \undefined \def \showCODEN     #1{\unskip}     \fi
\ifx \showDOI      \undefined \def \showDOI       #1{#1}\fi
\ifx \showISBNx    \undefined \def \showISBNx     #1{\unskip}     \fi
\ifx \showISBNxiii \undefined \def \showISBNxiii  #1{\unskip}     \fi
\ifx \showISSN     \undefined \def \showISSN      #1{\unskip}     \fi
\ifx \showLCCN     \undefined \def \showLCCN      #1{\unskip}     \fi
\ifx \shownote     \undefined \def \shownote      #1{#1}          \fi
\ifx \showarticletitle \undefined \def \showarticletitle #1{#1}   \fi
\ifx \showURL      \undefined \def \showURL       {\relax}        \fi
\providecommand\bibfield[2]{#2}
\providecommand\bibinfo[2]{#2}
\providecommand\natexlab[1]{#1}
\providecommand\showeprint[2][]{arXiv:#2}

\bibitem[Aaronson(2021)]%
        {Aaronson}
\bibfield{author}{\bibinfo{person}{Trevor Aaronson}.}
  \bibinfo{year}{2021}\natexlab{}.
\newblock \bibinfo{booktitle}{\emph{Spy in Disguise}}.
\newblock
\urldef\tempurl%
\url{https://theintercept.com/2021/09/12/fbi-informant-surveillance-muslims-supreme-court-911/}
\showURL{%
\tempurl}


\bibitem[Abadi(2015)]%
        {abadi2015majority}
\bibfield{author}{\bibinfo{person}{Mark Abadi}.}
  \bibinfo{year}{2015}\natexlab{}.
\newblock \showarticletitle{A majority of republicans in a new poll support
  Trump's proposal to bar Muslims from entering the US}.
\newblock \bibinfo{journal}{\emph{Business Insider. Retrieved from
  http://www.businessinsider.com/a-majority-of-republicans-support-trumps-proposal-to-ban-muslims-from-the-us-2015-12}}
  (\bibinfo{year}{2015}).
\newblock


\bibitem[Abu-Ras et~al\mbox{.}(2018)]%
        {abu2018muslim}
\bibfield{author}{\bibinfo{person}{Wahiba Abu-Ras}, \bibinfo{person}{Zulema~E
  Su{\'a}rez}, {and} \bibinfo{person}{Soleman Abu-Bader}.}
  \bibinfo{year}{2018}\natexlab{}.
\newblock \showarticletitle{Muslim Americans' safety and well-being in the wake
  of Trump: A public health and social justice crisis.}
\newblock \bibinfo{journal}{\emph{American Journal of Orthopsychiatry}}
  \bibinfo{volume}{88}, \bibinfo{number}{5} (\bibinfo{year}{2018}),
  \bibinfo{pages}{503}.
\newblock


\bibitem[Abugideiri(2010)]%
        {Abugideiri}
\bibfield{author}{\bibinfo{person}{Salma Elkadi Abugideiri}.}
  \bibinfo{year}{2010}\natexlab{}.
\newblock \showarticletitle{A Perspective on Domestic Violence in the Muslim
  Community}.
\newblock  (\bibinfo{year}{2010}).
\newblock
\urldef\tempurl%
\url{https://www.faithtrustinstitute.org/resources/articles/DV-in-Muslim-Community.pdf}
\showURL{%
\tempurl}


\bibitem[ACLU(2017)]%
        {ACLU}
\bibfield{author}{\bibinfo{person}{ACLU}.} \bibinfo{year}{2017}\natexlab{}.
\newblock \bibinfo{booktitle}{\emph{Anti-Muslim discrimination. American Civil
  Liberties Union.}}
\newblock
\urldef\tempurl%
\url{https://www.aclu.org/issues/national-security/discriminatory-profiling/anti-muslim-discrimination}
\showURL{%
\tempurl}


\bibitem[Afnan et~al\mbox{.}(2022)]%
        {afnan2022aunties}
\bibfield{author}{\bibinfo{person}{Tanisha Afnan}, \bibinfo{person}{Yixin Zou},
  \bibinfo{person}{Maryam Mustafa}, \bibinfo{person}{Mustafa Naseem}, {and}
  \bibinfo{person}{Florian Schaub}.} \bibinfo{year}{2022}\natexlab{}.
\newblock \showarticletitle{Aunties, Strangers, and the $\{$FBI$\}$: Online
  Privacy Concerns and Experiences of $\{$Muslim-American$\}$ Women}. In
  \bibinfo{booktitle}{\emph{Eighteenth Symposium on Usable Privacy and Security
  (SOUPS 2022)}}. \bibinfo{pages}{387--406}.
\newblock


\bibitem[Afrouz et~al\mbox{.}(2020)]%
        {afrouz2020seeking}
\bibfield{author}{\bibinfo{person}{Rojan Afrouz}, \bibinfo{person}{Beth~R
  Crisp}, {and} \bibinfo{person}{Ann Taket}.} \bibinfo{year}{2020}\natexlab{}.
\newblock \showarticletitle{Seeking help in domestic violence among Muslim
  women in Muslim-majority and non-Muslim-majority countries: A literature
  review}.
\newblock \bibinfo{journal}{\emph{Trauma, Violence, \& Abuse}}
  \bibinfo{volume}{21}, \bibinfo{number}{3} (\bibinfo{year}{2020}),
  \bibinfo{pages}{551--566}.
\newblock


\bibitem[Ali(2017)]%
        {ali2017impact}
\bibfield{author}{\bibinfo{person}{Areeza Ali}.}
  \bibinfo{year}{2017}\natexlab{}.
\newblock \showarticletitle{The impact of Islamophobia on the Muslim American
  community: accounts of psychological suffering, identity negotiation, and
  collective trauma}.
\newblock  (\bibinfo{year}{2017}).
\newblock


\bibitem[Alkhateeb(2012)]%
        {RN195}
\bibfield{author}{\bibinfo{person}{Maha Alkhateeb}.}
  \bibinfo{year}{2012}\natexlab{}.
\newblock \showarticletitle{Islamic Marriage Contracts}.
\newblock \bibinfo{journal}{\emph{Peaceful Families Project}}
  (\bibinfo{year}{2012}).
\newblock


\bibitem[Alkhateeb(nd)]%
        {mslmpowerctrl}
\bibfield{author}{\bibinfo{person}{Sharifa Alkhateeb}.}
  \bibinfo{year}{n.d.}\natexlab{}.
\newblock \bibinfo{booktitle}{\emph{Power and Control in Muslim Families}}.
\newblock
\urldef\tempurl%
\url{peacefulfamilies.org/uploads/1/1/0/5/110506531/power___control_wheel_for_muslim_families__1_.jpg}
\showURL{%
\tempurl}


\bibitem[Allen(2020)]%
        {allen2020towards}
\bibfield{author}{\bibinfo{person}{Chris Allen}.}
  \bibinfo{year}{2020}\natexlab{}.
\newblock \showarticletitle{Towards a Working Definition: Islamophobia and Its
  Contestation}.
\newblock In \bibinfo{booktitle}{\emph{Reconfiguring Islamophobia}}.
  \bibinfo{publisher}{Springer}, \bibinfo{pages}{1--13}.
\newblock


\bibitem[API-GBV(nd)]%
        {api}
\bibfield{author}{\bibinfo{person}{API-GBV}.} \bibinfo{year}{n.d.}\natexlab{}.
\newblock \showarticletitle{PATRIARCHY \& POWER}.
\newblock  (\bibinfo{year}{n.d.}).
\newblock
\urldef\tempurl%
\url{https://www.api-gbv.org/about-gbv/our-analysis/patriarchy-power/}
\showURL{%
\tempurl}


\bibitem[Arief et~al\mbox{.}(2014)]%
        {RN63}
\bibfield{author}{\bibinfo{person}{Budi Arief}, \bibinfo{person}{Kovila~P.L.
  Coopamootoo}, \bibinfo{person}{Martin Emms}, {and} \bibinfo{person}{Aad~van
  Moorsel}.} \bibinfo{year}{2014}\natexlab{}.
\newblock \bibinfo{title}{Sensible Privacy: How We Can Protect Domestic
  Violence Survivors Without Facilitating Misuse}.
\newblock , \bibinfo{numpages}{201-204}~pages.
\newblock
\urldef\tempurl%
\url{https://doi.org/10.1145/2665943.2665965}
\showDOI{\tempurl}


\bibitem[Armatta(2018)]%
        {armatta2018ending}
\bibfield{author}{\bibinfo{person}{Judith Armatta}.}
  \bibinfo{year}{2018}\natexlab{}.
\newblock \showarticletitle{Ending sexual violence through transformative
  justice}.
\newblock \bibinfo{journal}{\emph{Interdisciplinary Journal of Partnership
  Studies}} \bibinfo{volume}{5}, \bibinfo{number}{1} (\bibinfo{year}{2018}),
  \bibinfo{pages}{4--4}.
\newblock


\bibitem[Asad(2019)]%
        {Asad}
\bibfield{author}{\bibinfo{person}{Mariam Asad}.}
  \bibinfo{year}{2019}\natexlab{}.
\newblock \showarticletitle{Prefigurative Design as a Method for Research
  Justice}.
\newblock \bibinfo{journal}{\emph{Proc. ACM Hum.-Comput. Interact.}}
  \bibinfo{volume}{3}, \bibinfo{number}{CSCW}, Article \bibinfo{articleno}{200}
  (\bibinfo{date}{nov} \bibinfo{year}{2019}), \bibinfo{numpages}{18}~pages.
\newblock
\urldef\tempurl%
\url{https://doi.org/10.1145/3359302}
\showDOI{\tempurl}


\bibitem[Bassiouni(2012)]%
        {ummah}
\bibfield{author}{\bibinfo{person}{M.~Cherif Bassiouni}.}
  \bibinfo{year}{2012}\natexlab{}.
\newblock \bibinfo{booktitle}{\emph{The Social System and Morality of Islam}}.
\newblock
\urldef\tempurl%
\url{https://www.mei.edu/publications/social-system-and-morality-islam}
\showURL{%
\tempurl}


\bibitem[Bellini et~al\mbox{.}(2020b)]%
        {Bellini2020}
\bibfield{author}{\bibinfo{person}{Rosanna Bellini}, \bibinfo{person}{Simon
  Forrest}, \bibinfo{person}{Nicole Westmarland}, \bibinfo{person}{Dan
  Jackson}, {and} \bibinfo{person}{Jan~David Smeddinck}.}
  \bibinfo{year}{2020}\natexlab{b}.
\newblock \showarticletitle{Choice-Point: Fostering Awareness and Choice with
  Perpetrators in Domestic Violence Interventions}. In
  \bibinfo{booktitle}{\emph{Proceedings of the 2020 CHI Conference on Human
  Factors in Computing Systems}} (Honolulu, HI, USA)
  \emph{(\bibinfo{series}{CHI '20})}. \bibinfo{publisher}{Association for
  Computing Machinery}, \bibinfo{address}{New York, NY, USA},
  \bibinfo{pages}{1–14}.
\newblock
\showISBNx{9781450367080}
\urldef\tempurl%
\url{https://doi.org/10.1145/3313831.3376386}
\showDOI{\tempurl}


\bibitem[Bellini et~al\mbox{.}(2020a)]%
        {mechanisms}
\bibfield{author}{\bibinfo{person}{Rosanna Bellini}, \bibinfo{person}{Simon
  Forrest}, \bibinfo{person}{Nicole Westmarland}, {and}
  \bibinfo{person}{Jan~David Smeddinck}.} \bibinfo{year}{2020}\natexlab{a}.
\newblock \showarticletitle{Mechanisms of Moral Responsibility: Rethinking
  Technologies for Domestic Violence Prevention Work}.
\newblock  (\bibinfo{year}{2020}), \bibinfo{pages}{1–13}.
\newblock
\urldef\tempurl%
\url{https://doi.org/10.1145/3313831.3376693}
\showDOI{\tempurl}


\bibitem[Bellini et~al\mbox{.}(2019)]%
        {bellini2019mapping}
\bibfield{author}{\bibinfo{person}{Rosanna Bellini}, \bibinfo{person}{Angelika
  Strohmayer}, \bibinfo{person}{Patrick Olivier}, {and} \bibinfo{person}{Clara
  Crivellaro}.} \bibinfo{year}{2019}\natexlab{}.
\newblock \showarticletitle{Mapping the margins: Navigating the ecologies of
  domestic violence service provision}. In
  \bibinfo{booktitle}{\emph{Proceedings of the 2019 CHI Conference on Human
  Factors in Computing Systems}}. \bibinfo{pages}{1--13}.
\newblock


\bibitem[Bernstein and Gray(1997)]%
        {bernstein1997social}
\bibfield{author}{\bibinfo{person}{Andrea Bernstein} {and} \bibinfo{person}{Mel
  Gray}.} \bibinfo{year}{1997}\natexlab{}.
\newblock \bibinfo{booktitle}{\emph{Social work: a beginner's text}}.
\newblock \bibinfo{publisher}{Juta Legal and Academic Publishers}.
\newblock


\bibitem[Blackwell et~al\mbox{.}(2017)]%
        {blackwell2017classification}
\bibfield{author}{\bibinfo{person}{Lindsay Blackwell}, \bibinfo{person}{Jill
  Dimond}, \bibinfo{person}{Sarita Schoenebeck}, {and} \bibinfo{person}{Cliff
  Lampe}.} \bibinfo{year}{2017}\natexlab{}.
\newblock \showarticletitle{Classification and its consequences for online
  harassment: Design insights from heartmob}.
\newblock \bibinfo{journal}{\emph{Proceedings of the ACM on Human-Computer
  Interaction}} \bibinfo{volume}{1}, \bibinfo{number}{CSCW}
  (\bibinfo{year}{2017}), \bibinfo{pages}{1--19}.
\newblock


\bibitem[Brown et~al\mbox{.}(2014)]%
        {brown}
\bibfield{author}{\bibinfo{person}{Deana Brown}, \bibinfo{person}{Victoria
  Ayo}, {and} \bibinfo{person}{Rebecca~E. Grinter}.}
  \bibinfo{year}{2014}\natexlab{}.
\newblock \showarticletitle{Reflection through Design: Immigrant Women's
  Self-Reflection on Managing Health and Wellness}. In
  \bibinfo{booktitle}{\emph{Proceedings of the SIGCHI Conference on Human
  Factors in Computing Systems}} (Toronto, Ontario, Canada)
  \emph{(\bibinfo{series}{CHI '14})}. \bibinfo{publisher}{Association for
  Computing Machinery}, \bibinfo{address}{New York, NY, USA},
  \bibinfo{pages}{1605–1614}.
\newblock
\showISBNx{9781450324731}
\urldef\tempurl%
\url{https://doi.org/10.1145/2556288.2557119}
\showDOI{\tempurl}


\bibitem[Center(2015)]%
        {wilcenter}
\bibfield{author}{\bibinfo{person}{Wilson Center}.}
  \bibinfo{year}{2015}\natexlab{}.
\newblock \bibinfo{booktitle}{\emph{What is Gender-based Violence}}.
\newblock
\urldef\tempurl%
\url{https://gbv.wilsoncenter.org/what-gender-based-violence}
\showURL{%
\tempurl}


\bibitem[Chatterjee(2019)]%
        {chatterjee}
\bibfield{author}{\bibinfo{person}{Dom Chatterjee}.}
  \bibinfo{year}{2019}\natexlab{}.
\newblock \bibinfo{booktitle}{\emph{Centering survivors is not disposability
  culture: On who's responsible for transformative justice. Rest For
  Resistance.}}
\newblock


\bibitem[Chen et~al\mbox{.}(2022)]%
        {chen2022trauma}
\bibfield{author}{\bibinfo{person}{Janet~X Chen}, \bibinfo{person}{Allison
  McDonald}, \bibinfo{person}{Yixin Zou}, \bibinfo{person}{Emily Tseng},
  \bibinfo{person}{Kevin~A Roundy}, \bibinfo{person}{Acar Tamersoy},
  \bibinfo{person}{Florian Schaub}, \bibinfo{person}{Thomas Ristenpart}, {and}
  \bibinfo{person}{Nicola Dell}.} \bibinfo{year}{2022}\natexlab{}.
\newblock \showarticletitle{Trauma-Informed Computing: Towards Safer Technology
  Experiences for All}. In \bibinfo{booktitle}{\emph{CHI Conference on Human
  Factors in Computing Systems}}. \bibinfo{pages}{1--20}.
\newblock


\bibitem[Chordia(2022)]%
        {chordiaTJ}
\bibfield{author}{\bibinfo{person}{Ishita Chordia}.}
  \bibinfo{year}{2022}\natexlab{}.
\newblock \showarticletitle{Leveraging Transformative Justice in Organizing
  Collective Action Towards Community Safety}. In
  \bibinfo{booktitle}{\emph{Extended Abstracts of the 2022 CHI Conference on
  Human Factors in Computing Systems}} (New Orleans, LA, USA)
  \emph{(\bibinfo{series}{CHI EA '22})}. \bibinfo{publisher}{Association for
  Computing Machinery}, \bibinfo{address}{New York, NY, USA}, Article
  \bibinfo{articleno}{50}, \bibinfo{numpages}{4}~pages.
\newblock
\showISBNx{9781450391566}
\urldef\tempurl%
\url{https://doi.org/10.1145/3491101.3503820}
\showDOI{\tempurl}


\bibitem[Chou and Fu(2017)]%
        {chou2017influence}
\bibfield{author}{\bibinfo{person}{Li-Chen Chou} {and}
  \bibinfo{person}{Chung-Yuan Fu}.} \bibinfo{year}{2017}\natexlab{}.
\newblock \showarticletitle{The influence of Internet on politics: the impact
  of Facebook and the Internet penetration on elections in Taiwan}.
\newblock \bibinfo{journal}{\emph{Applied Economics Letters}}
  \bibinfo{volume}{24}, \bibinfo{number}{7} (\bibinfo{year}{2017}),
  \bibinfo{pages}{494--497}.
\newblock


\bibitem[Chrisler and Ferguson(2006)]%
        {chrisler2006violence}
\bibfield{author}{\bibinfo{person}{Joan~C Chrisler} {and}
  \bibinfo{person}{Sheila Ferguson}.} \bibinfo{year}{2006}\natexlab{}.
\newblock \showarticletitle{Violence against women as a public health issue}.
\newblock \bibinfo{journal}{\emph{Annals of the New York Academy of Sciences}}
  \bibinfo{volume}{1087}, \bibinfo{number}{1} (\bibinfo{year}{2006}),
  \bibinfo{pages}{235--249}.
\newblock


\bibitem[Chung(2010)]%
        {chung2010benefits}
\bibfield{author}{\bibinfo{person}{Jae~Eun Chung}.}
  \bibinfo{year}{2010}\natexlab{}.
\newblock \bibinfo{booktitle}{\emph{Benefits of social networking in online
  social support groups}}.
\newblock \bibinfo{publisher}{University of Southern California}.
\newblock


\bibitem[Clarke et~al\mbox{.}(2013)]%
        {RN67}
\bibfield{author}{\bibinfo{person}{Rachel Clarke}, \bibinfo{person}{Peter
  Wright}, \bibinfo{person}{Madeline Balaam}, {and} \bibinfo{person}{John
  McCarthy}.} \bibinfo{year}{2013}\natexlab{}.
\newblock \bibinfo{title}{Digital portraits: photo-sharing after domestic
  violence}.
\newblock , \bibinfo{numpages}{2517-2526}~pages.
\newblock
\urldef\tempurl%
\url{https://doi.org/10.1145/2470654.2481348}
\showDOI{\tempurl}


\bibitem[Coker(2002)]%
        {coker}
\bibfield{author}{\bibinfo{person}{Donna Coker}.}
  \bibinfo{year}{2002}\natexlab{}.
\newblock \showarticletitle{Transformative Justice: Anti-subordination
  Processes in Cases of Domestic Violence.}
\newblock
\urldef\tempurl%
\url{chrome-extension://efaidnbmnnnibpcajpcglclefindmkaj/https://repository.law.miami.edu/cgi/viewcontent.cgi?article=1229&context=fac_books}
\showURL{%
\tempurl}


\bibitem[Costanza-Chock(2018)]%
        {costanza2018design}
\bibfield{author}{\bibinfo{person}{Sasha Costanza-Chock}.}
  \bibinfo{year}{2018}\natexlab{}.
\newblock \showarticletitle{Design justice: Towards an intersectional feminist
  framework for design theory and practice}.
\newblock \bibinfo{journal}{\emph{Proceedings of the Design Research Society}}
  (\bibinfo{year}{2018}).
\newblock


\bibitem[Cross(2019)]%
        {harmreduction}
\bibfield{author}{\bibinfo{person}{Courtney Cross}.}
  \bibinfo{year}{2019}\natexlab{}.
\newblock \bibinfo{booktitle}{\emph{Harm Reduction in the Domestic Violence
  Context}}.
\newblock \bibinfo{pages}{332--361}.
\newblock
\showISBNx{9781479805648}
\urldef\tempurl%
\url{https://doi.org/10.18574/nyu/9781479805648.003.0014}
\showDOI{\tempurl}


\bibitem[Culture(nd)]%
        {forceVS}
\bibfield{author}{\bibinfo{person}{FORCE: Upsetting~Rape Culture}.}
  \bibinfo{year}{n.d.}\natexlab{}.
\newblock \bibinfo{booktitle}{\emph{Survivor, Victim, Victim-Survivor.}}
\newblock
\urldef\tempurl%
\url{https://upsettingrapeculture.com/survivor-victim}
\showURL{%
\tempurl}


\bibitem[Dieterle(2015)]%
        {dieterle2015designing}
\bibfield{author}{\bibinfo{person}{Brandy Dieterle}.}
  \bibinfo{year}{2015}\natexlab{}.
\newblock \showarticletitle{Designing smartphone apps for at risk populations:
  domestic violence survivors and user experience}. In
  \bibinfo{booktitle}{\emph{Proceedings of the 33rd annual international
  conference on the Design of Communication}}. \bibinfo{pages}{1--2}.
\newblock


\bibitem[Dombrowski et~al\mbox{.}(2016)]%
        {SJOR}
\bibfield{author}{\bibinfo{person}{Lynn Dombrowski}, \bibinfo{person}{Ellie
  Harmon}, {and} \bibinfo{person}{Sarah Fox}.} \bibinfo{year}{2016}\natexlab{}.
\newblock \bibinfo{title}{Social Justice-Oriented Interaction Design: Outlining
  Key Design Strategies and Commitments}.
\newblock , \bibinfo{numpages}{656-671}~pages.
\newblock
\urldef\tempurl%
\url{https://doi.org/10.1145/2901790.2901861}
\showDOI{\tempurl}


\bibitem[Douglas(2011)]%
        {douglas}
\bibfield{author}{\bibinfo{person}{Linda Douglas}.}
  \bibinfo{year}{2011}\natexlab{}.
\newblock \bibinfo{booktitle}{\emph{Harm reduction in the context of Domestic
  Violence Services.}}
\newblock
\urldef\tempurl%
\url{http://opendoorsnh.blogspot.com/2011/10/harm-reduction-in-context-of-domestic.html}
\showURL{%
\tempurl}


\bibitem[Drost et~al\mbox{.}(2015)]%
        {drost2015restorative}
\bibfield{author}{\bibinfo{person}{Lisanne Drost}, \bibinfo{person}{Birgitt
  Haller}, \bibinfo{person}{Veronika Hofinger}, \bibinfo{person}{Tinka van~der
  Kooij}, \bibinfo{person}{Katinka L{\"u}nnemann}, {and}
  \bibinfo{person}{Annemieke Wolthuis}.} \bibinfo{year}{2015}\natexlab{}.
\newblock \showarticletitle{Restorative justice in cases of domestic violence}.
\newblock \bibinfo{journal}{\emph{Best practice examples between increasing
  mutual understanding and awareness of specific protection needs}}
  (\bibinfo{year}{2015}).
\newblock


\bibitem[Erete et~al\mbox{.}(2021)]%
        {TJCS}
\bibfield{author}{\bibinfo{person}{Sheena Erete}, \bibinfo{person}{Karla
  Thomas}, \bibinfo{person}{Denise Nacu}, \bibinfo{person}{Jessa Dickinson},
  \bibinfo{person}{Naomi Thompson}, {and} \bibinfo{person}{Nichole Pinkard}.}
  \bibinfo{year}{2021}\natexlab{}.
\newblock \showarticletitle{Applying a Transformative Justice Approach to
  Encourage the Participation of Black and Latina Girls in Computing}.
\newblock \bibinfo{journal}{\emph{ACM Trans. Comput. Educ.}}
  \bibinfo{volume}{21}, \bibinfo{number}{4}, Article \bibinfo{articleno}{27}
  (\bibinfo{date}{oct} \bibinfo{year}{2021}), \bibinfo{numpages}{24}~pages.
\newblock
\urldef\tempurl%
\url{https://doi.org/10.1145/3451345}
\showDOI{\tempurl}


\bibitem[Evidence(2019)]%
        {accountability}
\bibfield{author}{\bibinfo{person}{Leaving Evidence}.}
  \bibinfo{year}{2019}\natexlab{}.
\newblock \bibinfo{booktitle}{\emph{The Four Parts of Accountability \& How To
  Give A Genuine Apology}}.
\newblock
\urldef\tempurl%
\url{https://leavingevidence.wordpress.com/2019/12/18/how-to-give-a-good-apology-part-1-the-four-parts-of-accountability/}
\showURL{%
\tempurl}


\bibitem[Fiesler(2019)]%
        {fiesler}
\bibfield{author}{\bibinfo{person}{Casey Fiesler}.}
  \bibinfo{year}{2019}\natexlab{}.
\newblock \bibinfo{booktitle}{\emph{Why (and how) academics should blog their
  papers}}.
\newblock
\urldef\tempurl%
\url{https://cfiesler.medium.com/why-and-how-academics-should-blog-their-papers-e6869559b8ea}
\showURL{%
\tempurl}


\bibitem[Filimowicz and Tzankova(2018)]%
        {filimowicz2018new}
\bibfield{author}{\bibinfo{person}{Michael Filimowicz} {and}
  \bibinfo{person}{Veronika Tzankova}.} \bibinfo{year}{2018}\natexlab{}.
\newblock \bibinfo{booktitle}{\emph{New directions in third wave human-computer
  interaction: volume 1-technologies}}.
\newblock \bibinfo{publisher}{Springer}.
\newblock


\bibitem[Five(2007)]%
        {genfive}
\bibfield{author}{\bibinfo{person}{Generation Five}.}
  \bibinfo{year}{2007}\natexlab{}.
\newblock \bibinfo{booktitle}{\emph{Toward Transformative Justice Guide: a
  Liberatory Approach to Child Sexual Abuse \& Others Forms of Intimate \&
  Community Violence}}.
\newblock
\urldef\tempurl%
\url{http://www.generationfive.org/wp-content/uploads/2013/07/G5_Toward_Transformative_Justice-Document.pdf}
\showURL{%
\tempurl}


\bibitem[for Constitutional~Rights(2019)]%
        {musban}
\bibfield{author}{\bibinfo{person}{Center for Constitutional~Rights}.}
  \bibinfo{year}{2019}\natexlab{}.
\newblock \bibinfo{booktitle}{\emph{The Muslim ban: Discriminatory impacts and
  lack of accountability.}}
\newblock
\urldef\tempurl%
\url{https://ccrjustice.org/home/get-involved/tools-resources/publications/muslim-ban-discriminatory-impacts-and-lack}
\showURL{%
\tempurl}


\bibitem[Freed et~al\mbox{.}(2019)]%
        {freed2019}
\bibfield{author}{\bibinfo{person}{Diana Freed}, \bibinfo{person}{Sam Havron},
  \bibinfo{person}{Emily Tseng}, \bibinfo{person}{Andrea Gallardo},
  \bibinfo{person}{Rahul Chatterjee}, \bibinfo{person}{Thomas Ristenpart},
  {and} \bibinfo{person}{Nicola Dell}.} \bibinfo{year}{2019}\natexlab{}.
\newblock \showarticletitle{"Is My Phone Hacked?" Analyzing Clinical Computer
  Security Interventions with Survivors of Intimate Partner Violence}.
\newblock \bibinfo{journal}{\emph{Proc. ACM Hum.-Comput. Interact.}}
  \bibinfo{volume}{3}, \bibinfo{number}{CSCW}, Article \bibinfo{articleno}{202}
  (\bibinfo{date}{nov} \bibinfo{year}{2019}), \bibinfo{numpages}{24}~pages.
\newblock
\urldef\tempurl%
\url{https://doi.org/10.1145/3359304}
\showDOI{\tempurl}


\bibitem[Freed et~al\mbox{.}(2018)]%
        {freed2018}
\bibfield{author}{\bibinfo{person}{Diana Freed}, \bibinfo{person}{Jackeline
  Palmer}, \bibinfo{person}{Diana Minchala}, \bibinfo{person}{Karen Levy},
  \bibinfo{person}{Thomas Ristenpart}, {and} \bibinfo{person}{Nicola Dell}.}
  \bibinfo{year}{2018}\natexlab{}.
\newblock \bibinfo{title}{“A Stalker's Paradise”: How Intimate Partner
  Abusers Exploit Technology}.
\newblock , \bibinfo{numpages}{13}~pages.
\newblock
\urldef\tempurl%
\url{https://doi.org/10.1145/3173574.3174241}
\showDOI{\tempurl}


\bibitem[Freed et~al\mbox{.}(2017)]%
        {freed2017digital}
\bibfield{author}{\bibinfo{person}{Diana Freed}, \bibinfo{person}{Jackeline
  Palmer}, \bibinfo{person}{Diana~Elizabeth Minchala}, \bibinfo{person}{Karen
  Levy}, \bibinfo{person}{Thomas Ristenpart}, {and} \bibinfo{person}{Nicola
  Dell}.} \bibinfo{year}{2017}\natexlab{}.
\newblock \showarticletitle{Digital technologies and intimate partner violence:
  A qualitative analysis with multiple stakeholders}.
\newblock \bibinfo{journal}{\emph{Proceedings of the ACM on human-computer
  interaction}} \bibinfo{volume}{1}, \bibinfo{number}{CSCW}
  (\bibinfo{year}{2017}), \bibinfo{pages}{1--22}.
\newblock


\bibitem[Freeland(2001)]%
        {freeland2001treatment}
\bibfield{author}{\bibinfo{person}{Richard Freeland}.}
  \bibinfo{year}{2001}\natexlab{}.
\newblock \showarticletitle{The treatment of Muslims in American courts}.
\newblock \bibinfo{journal}{\emph{Islam and Christian--Muslim Relations}}
  \bibinfo{volume}{12}, \bibinfo{number}{4} (\bibinfo{year}{2001}),
  \bibinfo{pages}{449--463}.
\newblock


\bibitem[Fujiwara et~al\mbox{.}(2021)]%
        {fujiwara2021effect}
\bibfield{author}{\bibinfo{person}{Thomas Fujiwara}, \bibinfo{person}{Karsten
  M{\"u}ller}, {and} \bibinfo{person}{Carlo Schwarz}.}
  \bibinfo{year}{2021}\natexlab{}.
\newblock \bibinfo{booktitle}{\emph{The effect of social media on elections:
  Evidence from the United States}}.
\newblock \bibinfo{type}{{T}echnical {R}eport}. \bibinfo{institution}{National
  Bureau of Economic Research}.
\newblock


\bibitem[Fusch Ph~D and Ness(2015)]%
        {fusch2015we}
\bibfield{author}{\bibinfo{person}{Patricia~I Fusch Ph~D} {and}
  \bibinfo{person}{Lawrence~R Ness}.} \bibinfo{year}{2015}\natexlab{}.
\newblock \showarticletitle{Are we there yet? Data saturation in qualitative
  research}.
\newblock  (\bibinfo{year}{2015}).
\newblock


\bibitem[Ghafournia(2017)]%
        {ghafournia2017muslim}
\bibfield{author}{\bibinfo{person}{Nafiseh Ghafournia}.}
  \bibinfo{year}{2017}\natexlab{}.
\newblock \showarticletitle{Muslim women and domestic violence: Developing a
  framework for social work practice}.
\newblock \bibinfo{journal}{\emph{Journal of Religion \& Spirituality in Social
  Work: Social Thought}} \bibinfo{volume}{36}, \bibinfo{number}{1-2}
  (\bibinfo{year}{2017}), \bibinfo{pages}{146--163}.
\newblock


\bibitem[Glickhouse(2017)]%
        {ProPublica}
\bibfield{author}{\bibinfo{person}{Rachel Glickhouse}.}
  \bibinfo{year}{2017}\natexlab{}.
\newblock \bibinfo{booktitle}{\emph{What we discovered during a year of
  documenting hate.}}
\newblock
\urldef\tempurl%
\url{https://www.propublica.org/article/what-we-discovered-during-a-year-of-documenting-hate}
\showURL{%
\tempurl}


\bibitem[G{\"o}kar{\i}ksel and Smith(2017)]%
        {gokariksel2017intersectional}
\bibfield{author}{\bibinfo{person}{Banu G{\"o}kar{\i}ksel} {and}
  \bibinfo{person}{Sara Smith}.} \bibinfo{year}{2017}\natexlab{}.
\newblock \showarticletitle{Intersectional feminism beyond US flag hijab and
  pussy hats in Trump's America}.
\newblock \bibinfo{journal}{\emph{Gender, Place \& Culture}}
  \bibinfo{volume}{24}, \bibinfo{number}{5} (\bibinfo{year}{2017}),
  \bibinfo{pages}{628--644}.
\newblock


\bibitem[Goodman(1961)]%
        {snowball}
\bibfield{author}{\bibinfo{person}{Leo~A Goodman}.}
  \bibinfo{year}{1961}\natexlab{}.
\newblock \showarticletitle{Snowball sampling}.
\newblock \bibinfo{journal}{\emph{The annals of mathematical statistics}}
  (\bibinfo{year}{1961}), \bibinfo{pages}{148--170}.
\newblock
\showISSN{0003-4851}


\bibitem[Goodmark(2004)]%
        {goodmark2004law}
\bibfield{author}{\bibinfo{person}{Leigh Goodmark}.}
  \bibinfo{year}{2004}\natexlab{}.
\newblock \showarticletitle{Law Is the Answer-Do We Know That for Sure:
  Questioning the Efficacy of Legal Interventions for Battered Women}.
\newblock \bibinfo{journal}{\emph{. Louis U. Pub. L. Rev.}}
  \bibinfo{volume}{23} (\bibinfo{year}{2004}), \bibinfo{pages}{7}.
\newblock


\bibitem[Gottlieb(2021)]%
        {gottlieb2021case}
\bibfield{author}{\bibinfo{person}{Mara Gottlieb}.}
  \bibinfo{year}{2021}\natexlab{}.
\newblock \showarticletitle{The case for a cultural humility framework in
  social work practice}.
\newblock \bibinfo{journal}{\emph{Journal of Ethnic \& Cultural Diversity in
  Social Work}} \bibinfo{volume}{30}, \bibinfo{number}{6}
  (\bibinfo{year}{2021}), \bibinfo{pages}{463--481}.
\newblock


\bibitem[Hammer(2019)]%
        {hammer2019peaceful}
\bibfield{author}{\bibinfo{person}{Juliane Hammer}.}
  \bibinfo{year}{2019}\natexlab{}.
\newblock \showarticletitle{Peaceful Families}.
\newblock In \bibinfo{booktitle}{\emph{Peaceful Families}}.
  \bibinfo{publisher}{Princeton University Press}.
\newblock


\bibitem[Havron et~al\mbox{.}(2019)]%
        {havron2019clinical}
\bibfield{author}{\bibinfo{person}{Sam Havron}, \bibinfo{person}{Diana Freed},
  \bibinfo{person}{Rahul Chatterjee}, \bibinfo{person}{Damon McCoy},
  \bibinfo{person}{Nicola Dell}, {and} \bibinfo{person}{Thomas Ristenpart}.}
  \bibinfo{year}{2019}\natexlab{}.
\newblock \showarticletitle{Clinical computer security for victims of intimate
  partner violence}. In \bibinfo{booktitle}{\emph{28th USENIX Security
  Symposium (USENIX Security 19)}}. \bibinfo{pages}{105--122}.
\newblock


\bibitem[Herz and Johansson(2012)]%
        {herz2012doing}
\bibfield{author}{\bibinfo{person}{Marcus Herz} {and} \bibinfo{person}{Thomas
  Johansson}.} \bibinfo{year}{2012}\natexlab{}.
\newblock \showarticletitle{‘Doing'social work: Critical considerations on
  theory and practice in social work}.
\newblock \bibinfo{journal}{\emph{Advances in social work}}
  \bibinfo{volume}{13}, \bibinfo{number}{3} (\bibinfo{year}{2012}),
  \bibinfo{pages}{527--540}.
\newblock


\bibitem[Hurley(2021)]%
        {pulse}
\bibfield{author}{\bibinfo{person}{Bevan Hurley}.}
  \bibinfo{year}{2021}\natexlab{}.
\newblock \bibinfo{booktitle}{\emph{Wife of Pulse nightclub shooter says she
  was subjected to years of abuse}}.
\newblock
\urldef\tempurl%
\url{https://www.independent.co.uk/news/world/americas/crime/pulse-nightclub-shooter-wife-abuse-b1938023.html}
\showURL{%
\tempurl}


\bibitem[II(2011)]%
        {TJhistory}
\bibfield{author}{\bibinfo{person}{Anthony J.~Nocella II}.}
  \bibinfo{year}{2011}\natexlab{}.
\newblock \showarticletitle{An Overview of the History and Theory of
  Transformative Justice}.
\newblock   \bibinfo{volume}{Volume 6 Issue 1} (\bibinfo{year}{2011}).
\newblock


\bibitem[Im et~al\mbox{.}(2021)]%
        {im2021yes}
\bibfield{author}{\bibinfo{person}{Jane Im}, \bibinfo{person}{Jill Dimond},
  \bibinfo{person}{Melody Berton}, \bibinfo{person}{Una Lee},
  \bibinfo{person}{Katherine Mustelier}, \bibinfo{person}{Mark~S Ackerman},
  {and} \bibinfo{person}{Eric Gilbert}.} \bibinfo{year}{2021}\natexlab{}.
\newblock \showarticletitle{Yes: Affirmative consent as a theoretical framework
  for understanding and imagining social platforms}. In
  \bibinfo{booktitle}{\emph{Proceedings of the 2021 CHI Conference on Human
  Factors in Computing Systems}}. \bibinfo{pages}{1--18}.
\newblock


\bibitem[Institute(2019)]%
        {islamo}
\bibfield{author}{\bibinfo{person}{Othering \&~Belonging Institute}.}
  \bibinfo{year}{2019}\natexlab{}.
\newblock \bibinfo{booktitle}{\emph{Consequences of Islamophobia on Civil
  Liberties and Rights in the United States.}}
\newblock
\urldef\tempurl%
\url{https://belonging.berkeley.edu/consequences-islamophobia-civil-liberties-and-rights-united-states}
\showURL{%
\tempurl}


\bibitem[Interventions(2012)]%
        {interventions2012creative}
\bibfield{author}{\bibinfo{person}{Creative Interventions}.}
  \bibinfo{year}{2012}\natexlab{}.
\newblock \bibinfo{title}{Creative Interventions Toolkit: A Practical Guide to
  Stop Interpersonal Violence. Pre-release version}.
\newblock
\newblock


\bibitem[Irani and Silberman(2013)]%
        {RN31}
\bibfield{author}{\bibinfo{person}{Lilly~C. Irani} {and}
  \bibinfo{person}{M.~Six Silberman}.} \bibinfo{year}{2013}\natexlab{}.
\newblock \bibinfo{title}{Turkopticon: interrupting worker invisibility in
  amazon mechanical turk}.
\newblock , \bibinfo{numpages}{611-620}~pages.
\newblock
\urldef\tempurl%
\url{https://doi.org/10.1145/2470654.2470742}
\showDOI{\tempurl}


\bibitem[Islam et~al\mbox{.}(2018)]%
        {islam2018challenges}
\bibfield{author}{\bibinfo{person}{Md~Jahirul Islam}, \bibinfo{person}{Masahiro
  Suzuki}, \bibinfo{person}{Nurunnahar Mazumder}, {and} \bibinfo{person}{Nada
  Ibrahim}.} \bibinfo{year}{2018}\natexlab{}.
\newblock \showarticletitle{Challenges of implementing restorative justice for
  intimate partner violence: An Islamic perspective}.
\newblock \bibinfo{journal}{\emph{Journal of Religion \& Spirituality in Social
  Work: Social Thought}} \bibinfo{volume}{37}, \bibinfo{number}{3}
  (\bibinfo{year}{2018}), \bibinfo{pages}{277--301}.
\newblock


\bibitem[Karamah(2015)]%
        {Karamah}
\bibfield{author}{\bibinfo{person}{Karamah}.} \bibinfo{year}{2015}\natexlab{}.
\newblock \bibinfo{booktitle}{\emph{Preventing \& responding to domestic
  violence: A workshop for Imams.}}
\newblock
\urldef\tempurl%
\url{https://karamah.org/ar/preventing-responding-
  to-domestic-violence-a-workshop-for-imams}
\showURL{%
\tempurl}


\bibitem[Kasturirangan et~al\mbox{.}(2004)]%
        {kasturirangan2004impact}
\bibfield{author}{\bibinfo{person}{Aarati Kasturirangan},
  \bibinfo{person}{Sandhya Krishnan}, {and} \bibinfo{person}{Stephanie Riger}.}
  \bibinfo{year}{2004}\natexlab{}.
\newblock \showarticletitle{The impact of culture and minority status on
  women's experience of domestic violence}.
\newblock \bibinfo{journal}{\emph{Trauma, Violence, \& Abuse}}
  \bibinfo{volume}{5}, \bibinfo{number}{4} (\bibinfo{year}{2004}),
  \bibinfo{pages}{318--332}.
\newblock


\bibitem[Kim(2010)]%
        {kim}
\bibfield{author}{\bibinfo{person}{Mimi Kim}.} \bibinfo{year}{2010}\natexlab{}.
\newblock \bibinfo{booktitle}{\emph{Innovative Strategies to Address Domestic
  Violence in Asian and Pacific Islander Communities: Examining Themes, Models
  and Interventions.}}
\newblock
\urldef\tempurl%
\url{https://www.api-gbv.org/resources/innovative-strategies/}
\showURL{%
\tempurl}


\bibitem[Kim(2021)]%
        {kim2021transformative}
\bibfield{author}{\bibinfo{person}{Mimi~E Kim}.}
  \bibinfo{year}{2021}\natexlab{}.
\newblock \showarticletitle{Transformative justice and restorative justice:
  Gender-based violence and alternative visions of justice in the United
  States}.
\newblock \bibinfo{journal}{\emph{International review of victimology}}
  \bibinfo{volume}{27}, \bibinfo{number}{2} (\bibinfo{year}{2021}),
  \bibinfo{pages}{162--172}.
\newblock


\bibitem[Kolko(2012)]%
        {wickedkolko}
\bibfield{author}{\bibinfo{person}{Jon Kolko}.}
  \bibinfo{year}{2012}\natexlab{}.
\newblock \bibinfo{booktitle}{\emph{Wicked problems: Problems worth solving}}.
\newblock \bibinfo{publisher}{Ac4d Austin, TX}.
\newblock
\showISBNx{0615593151}


\bibitem[Koyama(2001)]%
        {koyama2001toward}
\bibfield{author}{\bibinfo{person}{Emi Koyama}.}
  \bibinfo{year}{2001}\natexlab{}.
\newblock \showarticletitle{Toward a harm reduction approach in survivor
  advocacy}.
\newblock \bibinfo{journal}{\emph{Eminism. org}} (\bibinfo{year}{2001}).
\newblock


\bibitem[Kulkarni et~al\mbox{.}(2013)]%
        {kulkarni2013exploring}
\bibfield{author}{\bibinfo{person}{Shanti Kulkarni}, \bibinfo{person}{Holly
  Bell}, \bibinfo{person}{Jennifer~L Hartman}, {and} \bibinfo{person}{Robert~L
  Herman-Smith}.} \bibinfo{year}{2013}\natexlab{}.
\newblock \showarticletitle{Exploring individual and organizational factors
  contributing to compassion satisfaction, secondary traumatic stress, and
  burnout in domestic violence service providers}.
\newblock \bibinfo{journal}{\emph{Journal of the Society for Social Work and
  Research}} \bibinfo{volume}{4}, \bibinfo{number}{2} (\bibinfo{year}{2013}),
  \bibinfo{pages}{114--130}.
\newblock


\bibitem[Lajevardi and Oskooii(2018)]%
        {lajevardi2018old}
\bibfield{author}{\bibinfo{person}{Nazita Lajevardi} {and}
  \bibinfo{person}{Kassra~AR Oskooii}.} \bibinfo{year}{2018}\natexlab{}.
\newblock \showarticletitle{Old-fashioned racism, contemporary islamophobia,
  and the isolation of Muslim Americans in the age of Trump}.
\newblock \bibinfo{journal}{\emph{Journal of Race, Ethnicity, and Politics}}
  \bibinfo{volume}{3}, \bibinfo{number}{1} (\bibinfo{year}{2018}),
  \bibinfo{pages}{112--152}.
\newblock


\bibitem[Lee and Hadeed(2009)]%
        {lee2009intimate}
\bibfield{author}{\bibinfo{person}{Yeon-Shim Lee} {and} \bibinfo{person}{Linda
  Hadeed}.} \bibinfo{year}{2009}\natexlab{}.
\newblock \showarticletitle{Intimate partner violence among Asian immigrant
  communities: Health/mental health consequences, help-seeking behaviors, and
  service utilization}.
\newblock \bibinfo{journal}{\emph{Trauma, Violence, \& Abuse}}
  \bibinfo{volume}{10}, \bibinfo{number}{2} (\bibinfo{year}{2009}),
  \bibinfo{pages}{143--170}.
\newblock


\bibitem[Leit\~{a}o(2018)]%
        {Leitao}
\bibfield{author}{\bibinfo{person}{Roxanne Leit\~{a}o}.}
  \bibinfo{year}{2018}\natexlab{}.
\newblock \showarticletitle{Digital Technologies and Their Role in Intimate
  Partner Violence}. In \bibinfo{booktitle}{\emph{Extended Abstracts of the
  2018 CHI Conference on Human Factors in Computing Systems}} (Montreal QC,
  Canada) \emph{(\bibinfo{series}{CHI EA '18})}.
  \bibinfo{publisher}{Association for Computing Machinery},
  \bibinfo{address}{New York, NY, USA}, \bibinfo{pages}{1–6}.
\newblock
\showISBNx{9781450356213}
\urldef\tempurl%
\url{https://doi.org/10.1145/3170427.3180305}
\showDOI{\tempurl}


\bibitem[Lipka(2017)]%
        {lipka2017muslims}
\bibfield{author}{\bibinfo{person}{Michael Lipka}.}
  \bibinfo{year}{2017}\natexlab{}.
\newblock \showarticletitle{Muslims and Islam: Key findings in the US and
  around the world}.
\newblock  (\bibinfo{year}{2017}).
\newblock


\bibitem[Macfarlane(nd)]%
        {Macfarlane}
\bibfield{author}{\bibinfo{person}{Julie Macfarlane}.}
  \bibinfo{year}{n.d.}\natexlab{}.
\newblock \bibinfo{booktitle}{\emph{Family Dispute Processes Among North
  American Muslims}}.
\newblock
\urldef\tempurl%
\url{https://www.americanbar.org/groups/dispute_resolution/publications/dispute_resolution_magazine/2018/fall2018/family-dispute-processes-among-north-american-muslims/}
\showURL{%
\tempurl}


\bibitem[Maldonado and Murphy(2021)]%
        {maldonado2021does}
\bibfield{author}{\bibinfo{person}{Ana~I Maldonado} {and}
  \bibinfo{person}{Christopher~M Murphy}.} \bibinfo{year}{2021}\natexlab{}.
\newblock \showarticletitle{Does Trauma Help Explain the Need for Power and
  Control in Perpetrators of Intimate Partner Violence?}
\newblock \bibinfo{journal}{\emph{Journal of Family Violence}}
  \bibinfo{volume}{36}, \bibinfo{number}{3} (\bibinfo{year}{2021}),
  \bibinfo{pages}{347--359}.
\newblock


\bibitem[Matthews et~al\mbox{.}(2017)]%
        {RN8}
\bibfield{author}{\bibinfo{person}{Tara Matthews}, \bibinfo{person}{Kathleen
  O'Leary}, \bibinfo{person}{Anna Turner}, \bibinfo{person}{Manya Sleeper},
  \bibinfo{person}{Jill~Palzkill Woelfer}, \bibinfo{person}{Martin Shelton},
  \bibinfo{person}{Cori Manthorne}, \bibinfo{person}{Elizabeth F.~Churchill},
  {and} \bibinfo{person}{Sunny Consolvo}.} \bibinfo{year}{2017}\natexlab{}.
\newblock , \bibinfo{numpages}{2189-2201}~pages.
\newblock
\urldef\tempurl%
\url{https://doi.org/10.1145/3025453.3025875}
\showDOI{\tempurl}


\bibitem[Mieko~Yoshihama(2020)]%
        {Dabby}
\bibfield{author}{\bibinfo{person}{Shirley~Luo Mieko~Yoshihama, Chic~Dabby}.}
  \bibinfo{year}{2020}\natexlab{}.
\newblock \bibinfo{booktitle}{\emph{Facts \& Stats Report, Updated \& Expanded
  2020. Domestic Violence in Asian \& Pacific Islander Homes}}.
\newblock
\urldef\tempurl%
\url{https://www.api-gbv.org/resources/facts-stats-dv-api-homes/}
\showURL{%
\tempurl}


\bibitem[Mingus(2016)]%
        {pods}
\bibfield{author}{\bibinfo{person}{Mia Mingus}.}
  \bibinfo{year}{2016}\natexlab{}.
\newblock \bibinfo{booktitle}{\emph{Pods and Pod Mapping Worksheet}}.
\newblock
\urldef\tempurl%
\url{https://batjc.wordpress.com/resources/pods-and-pod-mapping-worksheet/}
\showURL{%
\tempurl}


\bibitem[Mingus(2019)]%
        {mia}
\bibfield{author}{\bibinfo{person}{Mia Mingus}.}
  \bibinfo{year}{2019}\natexlab{}.
\newblock \bibinfo{booktitle}{\emph{Transformative justice: A brief
  description.}}
\newblock
\urldef\tempurl%
\url{https://transformharm.org/transformative-justice-a-brief-description/}
\showURL{%
Retrieved August 22, 2020 from \tempurl}


\bibitem[Mohamed(2021)]%
        {pew2021}
\bibfield{author}{\bibinfo{person}{Besheer Mohamed}.}
  \bibinfo{year}{2021}\natexlab{}.
\newblock \bibinfo{booktitle}{\emph{Muslims are a growing presence in U.S., but
  still face negative views from the public}}.
\newblock
\urldef\tempurl%
\url{https://www.pewresearch.org/fact-tank/2021/09/01/muslims-are-a-growing-presence-in-u-s-but-still-face-negative-views-from-the-public/}
\showURL{%
\tempurl}


\bibitem[Ndjibu et~al\mbox{.}(2017)]%
        {ndjibu2017gender}
\bibfield{author}{\bibinfo{person}{Ruben Ndjibu}, \bibinfo{person}{Anicia~N
  Peters}, \bibinfo{person}{Heike Winschiers-Theophilus}, {and}
  \bibinfo{person}{Fannes Namhunya}.} \bibinfo{year}{2017}\natexlab{}.
\newblock \showarticletitle{Gender-based violence campaign in Namibia:
  traditional meets technology for societal change}. In
  \bibinfo{booktitle}{\emph{Proceedings of the 2017 CHI conference extended
  abstracts on human factors in computing systems}}.
  \bibinfo{pages}{1024--1029}.
\newblock


\bibitem[News and Media.(2020)]%
        {muslimapp}
\bibfield{author}{\bibinfo{person}{Guardian News} {and}
  \bibinfo{person}{Media.}} \bibinfo{year}{2020}\natexlab{}.
\newblock \bibinfo{booktitle}{\emph{ACLU files request over data us collected
  via Muslim app used by Millions.}}
\newblock
\urldef\tempurl%
\url{https://www.theguardian.com/us-news/2020/dec/03/aclu-seeks-release-records-data-us-collected-via-muslim-app-used-millions}
\showURL{%
\tempurl}


\bibitem[Nocella and Anthony(2011)]%
        {nocella2011overview}
\bibfield{author}{\bibinfo{person}{Anthony~J Nocella} {and} \bibinfo{person}{J
  Anthony}.} \bibinfo{year}{2011}\natexlab{}.
\newblock \showarticletitle{An overview of the history and theory of
  transformative justice}.
\newblock \bibinfo{journal}{\emph{Peace \& conflict review}}
  \bibinfo{volume}{6}, \bibinfo{number}{1} (\bibinfo{year}{2011}),
  \bibinfo{pages}{1--10}.
\newblock


\bibitem[of~Justice(2020)]%
        {USDOJ}
\bibfield{author}{\bibinfo{person}{US~Department of Justice}.}
  \bibinfo{year}{2020}\natexlab{}.
\newblock \showarticletitle{Federal Domestic Violence Laws}.
\newblock  (\bibinfo{year}{2020}).
\newblock
\urldef\tempurl%
\url{https://www.justice.gov/usao-wdtn/victim-witness-program/federal-domestic-violence-laws}
\showURL{%
\tempurl}


\bibitem[of~Social Workers~(NASW)(2021)]%
        {nasw}
\bibfield{author}{\bibinfo{person}{National~Association of Social
  Workers~(NASW)}.} \bibinfo{year}{2021}\natexlab{}.
\newblock \bibinfo{booktitle}{\emph{The Code of Ethics}}.
\newblock
\urldef\tempurl%
\url{https://www.socialworkers.org/About/Ethics/Code-of-Ethics/Code-of-Ethics-English}
\showURL{%
\tempurl}


\bibitem[Oyewuwo-Gassikia(2016)]%
        {oyewuwo2016american}
\bibfield{author}{\bibinfo{person}{Olubunmi~Basirat Oyewuwo-Gassikia}.}
  \bibinfo{year}{2016}\natexlab{}.
\newblock \showarticletitle{American Muslim women and domestic violence service
  seeking: A literature review}.
\newblock \bibinfo{journal}{\emph{Affilia}} \bibinfo{volume}{31},
  \bibinfo{number}{4} (\bibinfo{year}{2016}), \bibinfo{pages}{450--462}.
\newblock


\bibitem[Oyewuwo-Gassikia(2020)]%
        {oyewuwo2020black}
\bibfield{author}{\bibinfo{person}{Olubunmi~Basirat Oyewuwo-Gassikia}.}
  \bibinfo{year}{2020}\natexlab{}.
\newblock \showarticletitle{Black Muslim women's domestic violence help-seeking
  strategies: Types, motivations, and outcomes}.
\newblock \bibinfo{journal}{\emph{Journal of Aggression, Maltreatment \&
  Trauma}} \bibinfo{volume}{29}, \bibinfo{number}{7} (\bibinfo{year}{2020}),
  \bibinfo{pages}{856--875}.
\newblock


\bibitem[PFP(nd)]%
        {dvPFP}
\bibfield{author}{\bibinfo{person}{PFP}.} \bibinfo{year}{n.d.}\natexlab{}.
\newblock \bibinfo{booktitle}{\emph{About DV}}.
\newblock
\urldef\tempurl%
\url{https://www.peacefulfamilies.org/about-domestic-violence.html}
\showURL{%
\tempurl}


\bibitem[Rabaan(2021)]%
        {TJRabaan}
\bibfield{author}{\bibinfo{person}{Hawra Rabaan}.}
  \bibinfo{year}{2021}\natexlab{}.
\newblock \showarticletitle{Exploring Transformative Justice Principles to
  Inform Survivor-Centered Design for Muslim Women in the United States}. In
  \bibinfo{booktitle}{\emph{Companion Publication of the 2021 Conference on
  Computer Supported Cooperative Work and Social Computing}} (Virtual Event,
  USA) \emph{(\bibinfo{series}{CSCW '21})}. \bibinfo{publisher}{Association for
  Computing Machinery}, \bibinfo{address}{New York, NY, USA},
  \bibinfo{pages}{291–294}.
\newblock
\showISBNx{9781450384797}
\urldef\tempurl%
\url{https://doi.org/10.1145/3462204.3481797}
\showDOI{\tempurl}


\bibitem[Rabaan et~al\mbox{.}(2021)]%
        {rab21}
\bibfield{author}{\bibinfo{person}{Hawra Rabaan}, \bibinfo{person}{Alyson~L.
  Young}, {and} \bibinfo{person}{Lynn Dombrowski}.}
  \bibinfo{year}{2021}\natexlab{}.
\newblock \showarticletitle{Daughters of Men: Saudi Women's Sociotechnical
  Agency Practices in Addressing Domestic Abuse}.
\newblock \bibinfo{journal}{\emph{Proc. ACM Hum.-Comput. Interact.}}
  \bibinfo{volume}{4}, \bibinfo{number}{CSCW3}, Article
  \bibinfo{articleno}{224} (\bibinfo{date}{jan} \bibinfo{year}{2021}),
  \bibinfo{numpages}{31}~pages.
\newblock
\urldef\tempurl%
\url{https://doi.org/10.1145/3432923}
\showDOI{\tempurl}


\bibitem[Rana and Marin(2012)]%
        {rana2012addressing}
\bibfield{author}{\bibinfo{person}{Sheetal Rana} {and} \bibinfo{person}{L
  Marin}.} \bibinfo{year}{2012}\natexlab{}.
\newblock \showarticletitle{Addressing domestic violence in immigrant
  communities: Critical issues for culturally competent services}.
\newblock \bibinfo{journal}{\emph{Harrisburg, PA: VAWnet, a project of the
  National Resource Center on Domestic Violence}} (\bibinfo{year}{2012}).
\newblock


\bibitem[Rankine et~al\mbox{.}(2017)]%
        {rankine2017pacific}
\bibfield{author}{\bibinfo{person}{Jenny Rankine}, \bibinfo{person}{Teuila
  Percival}, \bibinfo{person}{Eseta Finau}, \bibinfo{person}{Linda-Teleo Hope},
  \bibinfo{person}{Pefi Kingi}, \bibinfo{person}{Maiava~Carmel Peteru},
  \bibinfo{person}{Elizabeth Powell}, \bibinfo{person}{Robert Robati-Mani},
  {and} \bibinfo{person}{Elisala Selu}.} \bibinfo{year}{2017}\natexlab{}.
\newblock \showarticletitle{Pacific peoples, violence, and the power and
  control wheel}.
\newblock \bibinfo{journal}{\emph{Journal of interpersonal violence}}
  \bibinfo{volume}{32}, \bibinfo{number}{18} (\bibinfo{year}{2017}),
  \bibinfo{pages}{2777--2803}.
\newblock


\bibitem[Rees and Pease(2007)]%
        {rees2007domestic}
\bibfield{author}{\bibinfo{person}{Susan Rees} {and} \bibinfo{person}{Bob
  Pease}.} \bibinfo{year}{2007}\natexlab{}.
\newblock \showarticletitle{Domestic violence in refugee families in
  Australia}.
\newblock \bibinfo{journal}{\emph{Journal of immigrant \& refugee studies}}
  \bibinfo{volume}{5}, \bibinfo{number}{2} (\bibinfo{year}{2007}),
  \bibinfo{pages}{1--19}.
\newblock


\bibitem[Rising(nd)]%
        {undoc}
\bibfield{author}{\bibinfo{person}{Immigrants Rising}.}
  \bibinfo{year}{n.d.}\natexlab{}.
\newblock \bibinfo{booktitle}{\emph{Defining Undocumented}}.
\newblock
\urldef\tempurl%
\url{https://immigrantsrising.org/resource/defining-undocumented/}
\showURL{%
\tempurl}


\bibitem[Said(2010)]%
        {said2010terrorist}
\bibfield{author}{\bibinfo{person}{Wadie~E Said}.}
  \bibinfo{year}{2010}\natexlab{}.
\newblock \showarticletitle{The terrorist informant}.
\newblock \bibinfo{journal}{\emph{Wash. L. Rev.}}  \bibinfo{volume}{85}
  (\bibinfo{year}{2010}), \bibinfo{pages}{687}.
\newblock


\bibitem[SAKI(2015)]%
        {SAKI}
\bibfield{author}{\bibinfo{person}{SAKI}.} \bibinfo{year}{2015}\natexlab{}.
\newblock \bibinfo{booktitle}{\emph{Victim or Survivor: Terminology Through
  Investigation to Prosecution.}}
\newblock
\urldef\tempurl%
\url{https://www.sakitta.org/toolkit/index.cfm?fuseaction=tool&tool=80}
\showURL{%
\tempurl}


\bibitem[SAMHSA(2014)]%
        {TIP}
\bibfield{author}{\bibinfo{person}{SAMHSA}.} \bibinfo{year}{2014}\natexlab{}.
\newblock \bibinfo{booktitle}{\emph{TIP57: Trauma-Informed Care in Behavioral
  Health Services}}.
\newblock
\urldef\tempurl%
\url{https://store.samhsa.gov/sites/default/files/d7/priv/sma14-4816.pdf}
\showURL{%
\tempurl}


\bibitem[Schoenebeck et~al\mbox{.}(2021)]%
        {schoenebeck2021drawing}
\bibfield{author}{\bibinfo{person}{Sarita Schoenebeck},
  \bibinfo{person}{Oliver~L Haimson}, {and} \bibinfo{person}{Lisa Nakamura}.}
  \bibinfo{year}{2021}\natexlab{}.
\newblock \showarticletitle{Drawing from justice theories to support targets of
  online harassment}.
\newblock \bibinfo{journal}{\emph{new media \& society}} \bibinfo{volume}{23},
  \bibinfo{number}{5} (\bibinfo{year}{2021}), \bibinfo{pages}{1278--1300}.
\newblock


\bibitem[Sedgwick(2000)]%
        {sedgwick2000sects}
\bibfield{author}{\bibinfo{person}{Mark Sedgwick}.}
  \bibinfo{year}{2000}\natexlab{}.
\newblock \showarticletitle{Sects in the Islamic World1}.
\newblock \bibinfo{journal}{\emph{Nova Religio}} \bibinfo{volume}{3},
  \bibinfo{number}{2} (\bibinfo{year}{2000}), \bibinfo{pages}{195--240}.
\newblock


\bibitem[Shiekh(2011)]%
        {shiekh2011detained}
\bibfield{author}{\bibinfo{person}{Irum Shiekh}.}
  \bibinfo{year}{2011}\natexlab{}.
\newblock \bibinfo{booktitle}{\emph{Detained Without Cause: Muslims' Stories of
  Detention and Deportation in America After 9/11}}.
\newblock \bibinfo{publisher}{Springer}.
\newblock


\bibitem[Sokoloff and Dupont(2005)]%
        {sokoloff2005domestic}
\bibfield{author}{\bibinfo{person}{Natalie~J Sokoloff} {and}
  \bibinfo{person}{Ida Dupont}.} \bibinfo{year}{2005}\natexlab{}.
\newblock \showarticletitle{Domestic violence at the intersections of race,
  class, and gender: Challenges and contributions to understanding violence
  against marginalized women in diverse communities}.
\newblock \bibinfo{journal}{\emph{Violence against women}}
  \bibinfo{volume}{11}, \bibinfo{number}{1} (\bibinfo{year}{2005}),
  \bibinfo{pages}{38--64}.
\newblock


\bibitem[Sterling(2013)]%
        {RN68}
\bibfield{author}{\bibinfo{person}{S.~Revi Sterling}.}
  \bibinfo{year}{2013}\natexlab{}.
\newblock \bibinfo{title}{Designing for trauma: the roles of ICTD in combating
  violence against women (VAW)}.
\newblock , \bibinfo{numpages}{159-162}~pages.
\newblock
\urldef\tempurl%
\url{https://doi.org/10.1145/2517899.2517908}
\showDOI{\tempurl}


\bibitem[Strauss and Corbin(1997)]%
        {strauss1997grounded}
\bibfield{author}{\bibinfo{person}{Anselm Strauss} {and}
  \bibinfo{person}{Juliet~M Corbin}.} \bibinfo{year}{1997}\natexlab{}.
\newblock \bibinfo{booktitle}{\emph{Grounded theory in practice}}.
\newblock \bibinfo{publisher}{Sage}.
\newblock


\bibitem[Stubbs(2002)]%
        {stubbs2002domestic}
\bibfield{author}{\bibinfo{person}{Julie Stubbs}.}
  \bibinfo{year}{2002}\natexlab{}.
\newblock \bibinfo{title}{Restorative justice and family violence. H. Strang
  and J. Braithwaite}.
\newblock
\newblock


\bibitem[Sultana et~al\mbox{.}(2018)]%
        {RN5}
\bibfield{author}{\bibinfo{person}{Sharifa Sultana}, \bibinfo{person}{Franois
  Guimbretiere}, \bibinfo{person}{Phoebe Sengers}, {and}
  \bibinfo{person}{Nicola Dell}.} \bibinfo{year}{2018}\natexlab{}.
\newblock \bibinfo{title}{Design Within a Patriarchal Society: Opportunities
  and Challenges in Designing for Rural Women in Bangladesh}.
\newblock , \bibinfo{numpages}{13}~pages.
\newblock
\urldef\tempurl%
\url{https://doi.org/10.1145/3173574.3174110}
\showDOI{\tempurl}


\bibitem[Sultana et~al\mbox{.}(2022)]%
        {TJsharifa}
\bibfield{author}{\bibinfo{person}{Sharifa Sultana},
  \bibinfo{person}{Sadia~Tasnuva Pritha}, \bibinfo{person}{Rahnuma Tasnim},
  \bibinfo{person}{Anik Das}, \bibinfo{person}{Rokeya Akter},
  \bibinfo{person}{Shaid Hasan}, \bibinfo{person}{S.M.~Raihanul Alam},
  \bibinfo{person}{Muhammad~Ashad Kabir}, {and} \bibinfo{person}{Syed~Ishtiaque
  Ahmed}.} \bibinfo{year}{2022}\natexlab{}.
\newblock \showarticletitle{‘ShishuShurokkha': A Transformative Justice
  Approach for Combating Child Sexual Abuse in Bangladesh}. In
  \bibinfo{booktitle}{\emph{Proceedings of the 2022 CHI Conference on Human
  Factors in Computing Systems}} (New Orleans, LA, USA)
  \emph{(\bibinfo{series}{CHI '22})}. \bibinfo{publisher}{Association for
  Computing Machinery}, \bibinfo{address}{New York, NY, USA}, Article
  \bibinfo{articleno}{577}, \bibinfo{numpages}{23}~pages.
\newblock
\showISBNx{9781450391573}
\urldef\tempurl%
\url{https://doi.org/10.1145/3491102.3517543}
\showDOI{\tempurl}


\bibitem[Tseng et~al\mbox{.}(2020)]%
        {tseng2020tools}
\bibfield{author}{\bibinfo{person}{Emily Tseng}, \bibinfo{person}{Rosanna
  Bellini}, \bibinfo{person}{Nora McDonald}, \bibinfo{person}{Matan Danos},
  \bibinfo{person}{Rachel Greenstadt}, \bibinfo{person}{Damon McCoy},
  \bibinfo{person}{Nicola Dell}, {and} \bibinfo{person}{Thomas Ristenpart}.}
  \bibinfo{year}{2020}\natexlab{}.
\newblock \showarticletitle{The tools and tactics used in intimate partner
  surveillance: An analysis of online infidelity forums}. In
  \bibinfo{booktitle}{\emph{29th USENIX Security Symposium (USENIX Security
  20)}}. \bibinfo{pages}{1893--1909}.
\newblock


\bibitem[Tseng et~al\mbox{.}(2022)]%
        {tseng2022care}
\bibfield{author}{\bibinfo{person}{Emily Tseng}, \bibinfo{person}{Mehrnaz
  Sabet}, \bibinfo{person}{Rosanna Bellini}, \bibinfo{person}{Harkiran~Kaur
  Sodhi}, \bibinfo{person}{Thomas Ristenpart}, {and} \bibinfo{person}{Nicola
  Dell}.} \bibinfo{year}{2022}\natexlab{}.
\newblock \showarticletitle{Care Infrastructures for Digital Security in
  Intimate Partner Violence}. In \bibinfo{booktitle}{\emph{CHI Conference on
  Human Factors in Computing Systems}}. \bibinfo{pages}{1--20}.
\newblock


\bibitem[Van~der Kolk(2022)]%
        {van2022posttraumatic}
\bibfield{author}{\bibinfo{person}{Bessel Van~der Kolk}.}
  \bibinfo{year}{2022}\natexlab{}.
\newblock \showarticletitle{Posttraumatic stress disorder and the nature of
  trauma}.
\newblock \bibinfo{journal}{\emph{Dialogues in clinical neuroscience}}
  (\bibinfo{year}{2022}).
\newblock


\bibitem[Verkerk(1999)]%
        {verkerk1999care}
\bibfield{author}{\bibinfo{person}{Marian Verkerk}.}
  \bibinfo{year}{1999}\natexlab{}.
\newblock \showarticletitle{A care perspective on coercian and autonomy}.
\newblock \bibinfo{journal}{\emph{Bioethics}} \bibinfo{volume}{13},
  \bibinfo{number}{3-4} (\bibinfo{year}{1999}), \bibinfo{pages}{358--368}.
\newblock


\bibitem[Violence.(nd)]%
        {NCDV}
\bibfield{author}{\bibinfo{person}{National Coalition Against~Domestic
  Violence.}} \bibinfo{year}{n.d.}\natexlab{}.
\newblock \bibinfo{booktitle}{\emph{National Statistics Domestic Violence Fact
  Sheet}}.
\newblock
\urldef\tempurl%
\url{https://ncadv.org/statistics}
\showURL{%
\tempurl}


\bibitem[Women(nd)]%
        {UNwomen}
\bibfield{author}{\bibinfo{person}{UN Women}.} \bibinfo{year}{n.d.}\natexlab{}.
\newblock \bibinfo{booktitle}{\emph{Self-Learning Booklet: Understanding
  Masculinities and Violence Against Women and Girls}}.
\newblock
\urldef\tempurl%
\url{https://www.fsnnetwork.org/resource/self-learning-booklet-understanding-masculinities-and-violence-against-women-and-girls}
\showURL{%
\tempurl}


\bibitem[Xiao et~al\mbox{.}(2022)]%
        {Xiao2022}
\bibfield{author}{\bibinfo{person}{Sijia Xiao}, \bibinfo{person}{Coye
  Cheshire}, {and} \bibinfo{person}{Niloufar Salehi}.}
  \bibinfo{year}{2022}\natexlab{}.
\newblock \showarticletitle{Sensemaking, Support, Safety, Retribution,
  Transformation: A Restorative Justice Approach to Understanding Adolescents'
  Needs for Addressing Online Harm}. In \bibinfo{booktitle}{\emph{Proceedings
  of the 2022 CHI Conference on Human Factors in Computing Systems}} (New
  Orleans, LA, USA) \emph{(\bibinfo{series}{CHI '22})}.
  \bibinfo{publisher}{Association for Computing Machinery},
  \bibinfo{address}{New York, NY, USA}, Article \bibinfo{articleno}{146},
  \bibinfo{numpages}{15}~pages.
\newblock
\showISBNx{9781450391573}
\urldef\tempurl%
\url{https://doi.org/10.1145/3491102.3517614}
\showDOI{\tempurl}


\bibitem[Zehr(2015)]%
        {zehr2015little}
\bibfield{author}{\bibinfo{person}{Howard Zehr}.}
  \bibinfo{year}{2015}\natexlab{}.
\newblock \bibinfo{booktitle}{\emph{The little book of restorative justice:
  Revised and updated}}.
\newblock \bibinfo{publisher}{Simon and Schuster}.
\newblock


\end{thebibliography}







\end{document}